\documentclass[a4paper,11pt]{article}
\pdfoutput=1 
\usepackage{jcappub} 
\usepackage[T1]{fontenc}

\newcommand{\gsim}{\raisebox{-0.13cm}{~\shortstack{$>$ \\[-0.07cm]
      $\sim$}}~}
\newcommand{\lsim}{\raisebox{-0.13cm}{~\shortstack{$<$ \\[-0.07cm]
      $\sim$}}~}

\usepackage{subfig}
\usepackage{comment}
\usepackage{graphicx}
\usepackage{amsmath}
\usepackage{amssymb}
\usepackage{lipsum}
\usepackage{nicefrac}
\usepackage{multirow}
\usepackage{xcolor}
\usepackage{hyperref}
\usepackage{natbib} 
\usepackage{multicol}
\usepackage{float}
\usepackage{enumitem}
\usepackage{soul}
\usepackage{subcaption}
\usepackage{lscape}
\usepackage[normalem]{ulem}
\usepackage[export]{adjustbox} 
\newcommand{\Tstrut}{\rule{0pt}{2.6ex}}  
\usepackage{aas_macros}

\defcitealias{Mead_2021}{M20}
\defcitealias{Takahashi_2012}{TB}
\defcitealias{Carbone_2016}{DEMNUni}

\title{\centering{\boldmath{Detecting the neutrino mass via the cross-correlation between matter tracers and the ISWRS effect?}}}

\author[a,b]{Viviana Cuozzo,}
\author[a,b]{Marina Migliaccio,}
\author[c]{Matteo Calabrese,}
\author[d]{Carmelita Carbone}

\affiliation[a]{Università degli Studi di Roma Tor Vergata, via della Ricerca Scientifica 1, I-00133 Roma, Italy}
\affiliation[b]{INFN, Sezione di Roma 2, Università degli Studi di Roma Tor Vergata, via della Ricerca Scientifica 1, I-00133 Roma, Italy }
\affiliation[d]{INAF -- Istituto di Astrofisica Spaziale e Fisica cosmica di Milano (IASF-MI), via Alfonso Corti 12, I-20133 Milano, Italy}
\affiliation[c]{Astronomical Observatory of the Autonomous Region of the Aosta Valley (OAVdA), Loc. Lignan 39, I-11020, Nus (Aosta Valley), Italy}
\affiliation[d]{INAF -- Istituto di Astrofisica Spaziale e Fisica cosmica di Milano (IASF-MI), via Alfonso Corti 12, I-20133 Milano, Italy}
\emailAdd{vcuozzo@roma2.infn.it, marina.migliaccio@roma2.infn.it, calabrese@oavda.it, carmelita.carbone@inaf.it}

\abstract{This work explores the potential to detect the nonlinear Integrated Sachs Wolfe effect, namely the Rees-Sciama effect (ISWRS), by cross-correlating current and future Cosmic Microwave Background (CMB) experiments—Simons Observatory, CMB-S4, CMB-HD, and PICO—with ongoing Large Scale Structure (LSS) surveys, such as Euclid and the Vera Rubin Observatory (LSST). We model the cross-correlation of the ISWRS effect with gravitational potential tracers like galaxy clustering, cosmic shear, and CMB-lensing potential, to forecast results from these experiments. Our analysis also accounts for the presence of massive neutrinos to assess the feasibility of identifying the $\nu\Lambda$CDM model and constraining the neutrino mass sum, $M_{\nu}$. Our findings indicate that the CMB-lensing potential reconstructed by CMB-HD is expected to provide the most promising results, achieving $\gsim5\sigma$ detections even under conservative assumptions for detector noise and foregrounds, thereby allowing differentiation between $\nu\Lambda$CDM models. 
Galaxy clustering can also yield significant detections, whereas cosmic shear can provide valuable results only if non-linearities are accurately modelled, beyond the capabilities of currently available analytical approaches.
These latter LSS probes do not provide strong constraining power on $M_{\nu}$. While our findings suggest that future CMB experiments and LSS surveys will enable the detection of the ISWRS effect, they do not imply significant prospects for imposing new constraints on neutrino masses in the near future.}

\begin{document}

\maketitle 

\section{Introduction}
\label{sec:intro}
Modern cosmology deals with several open questions. First among the others, the nature of Dark Energy (DE; e.g., \cite{Linder_2003}). Some of the unresolved issues are intertwined with the Standard Model of particle physics, such as the value of the sum of the three neutrino masses, $M_{\nu} = \sum m_{\nu}$, whose tightest bounds come in fact from cosmological data~\cite{Lesgourgues_2012, Arcidiacono_2020, PlanckVI_2020, Allali_2024, 2025Elbers_DESI, Jiang_2025}. The Integrated Sachs-Wolfe effect~\cite{Sachs_1967} (ISW), along with its nonlinear counterpart the Rees-Sciama effect~\cite{Rees_1968} (RS), can provide a means to explore both of these open issues. This is because the ISWRS effect consists of the temperature variation experienced by Cosmic Microwave Background (CMB) photons~\cite{Penzias_1965} when passing through a time-varying gravitational potential $\Phi$. In the linear regime, we can observe a non-zero $\dot\Phi$ either because of the accelerated expansion of the Universe, mainly induced by DE~\cite{Ferraro_2015, Giannantonio_2008, Giannantonio_2012, Watson_2014, Naidoo_2021} or modified gravity effects~\cite{Kimura_2012, Nakamura_2019, Kable_2022, SmithT_2023}, or because of spatial curvature~\cite{Kamionkowski}. In the nonlinear regime, $\Phi$ evolves primarily because of structure formation, and this is where neutrinos play a significant role. The presence of massive neutrinos induces in fact a slow decay of the gravitational potential~\cite{Bond_1980} that generates ISWRS even in the absence of a background expansion. Therefore, a full reconstruction of the ISWRS effect could provide new insight into the physics of DE and neutrinos, aiding in the constraints on the DE Equation of State (EoS) and on the total neutrino mass $M_{\nu}$. 

The ISWRS signal is, however, exceedingly faint if compared to the primary anisotropies of the CMB, and direct measurements have yet to be achieved. However, we can reconstruct this signal by exploiting its cross-correlation with cosmological probes that trace the gravitational potential, whose variations induce the ISWRS effect itself~\cite{Crittenden_1996, Kneissl_1997, Boughn_1998, Boughn_2001, Boughn_2004, Giannantonio_2008, Lesgourgues_2008, Douspis_2008}. Moreover, these correlations are strongly influenced by the presence of massive neutrinos~\cite{Cai_2010, Nishizawa_2014, Carbone_2016, Cuozzo_2024}, making their detection a potential avenue for addressing questions on $M_{\nu}$. Although cross-correlation with $\Phi$-tracers improves the chances of detection by helping isolate the ISWRS signal that would otherwise be buried in primary CMB anisotropies, foregrounds, and noise, it comes at the cost of adding the associated complexities of the $\Phi$-tracers to those of the CMB observations. The former generation of CMB experiments and Large Scale Structure (LSS) surveys has not been able to fully overcome the main obstacles that characterise this detection: cosmic variance limit, extragalactic foregrounds in CMB anisotropy maps, and the lack of data in the nonlinear regime, which is where neutrinos effects appear. However, ongoing advancements in CMB delensing~\cite{Hotinli_2022} and foreground removal techniques (e.g.~\cite{Puglisi_2022, 2025ACT_fg, 2020ACT_fg, 2025SPT_fg}) 
for CMB anisotropy maps, the implementation of analytical techniques to maximise the cross-correlation between CMB and LSS tracers (as studied in~\cite{Ferraro_2022}), combined with high-sensitivity and high-resolution measurements by upcoming experiments, can pave the way not only for the detection of the ISWRS effect but also, in principle, for distinguishing between different $M_{\nu}$ values.

In this work, we forecast the feasibility of both these achievements via the combination of current and next-generation CMB experiments, the Simons Observatory (SO)~\cite{SO_2019,Lee_2019}, a CMB-S4-like experiment~\cite{CMB-S4, S4_2016}, CMB-HD~\cite{HD_2019, CMBHD_2022} and the space mission PICO~\cite{Hanany_2019, Aurlien_2023}, with current  LSS surveys, such as Euclid~\cite{Euclid_24} and the Vera Rubin Observatory (hereafter, LSST)~\cite{LSST_2009,LSST_2018}. To perform this analysis we exploit the analytical method developed in~\cite{Cuozzo_2024}, to compute the cross-angular power spectra between the ISWRS effect and three $\Phi$-tracer probes: galaxy clustering (GC), cosmic shear (CS) and CMB-lensing potential (CMBL). Because of the different effect that different values of $M_{\nu}$ have on the cross-spectra, we explore five $\nu\Lambda$CDM cosmologies in our analysis, using as baseline the Planck 2018 one~\cite{PlanckVI_2020}. These studies are particularly relevant not only to produce forecasts for the considered ongoing and forthcoming experiments, but also because testing diverse experiments (i.e., ground-based and satellites) and under different conditions (i.e., different levels of delensing and foreground cleaning) enables to identify the optimal conditions for successfully detecting the ISWRS and, potentially, constraining $M_{\nu}$ in the future.

This paper is organised as follows. In Section~\ref{sec:theory} we summarise the theoretical framework for the ISWRS effect and its cross-correlation with tracers of the gravitational potential in the presence of massive neutrinos. In Section~\ref{sec:results} we present our findings on the detectability of the ISWRS effect through its cross-correlation with $\Phi$ tracers, for both ideal conditions and realistic scenarios. Additionally, we assess the potential of distinguishing between different $\nu\Lambda$CDM models through this detection. Finally, in Section~\ref{sec:conclusions} we draw our conclusions.

\section{Theoretical Framework}
\label{sec:theory}
\subsection{Integrated Sachs Wolfe and Rees Sciama Effects}
As a secondary anisotropy of the CMB, the ISW effect consists of a temperature variation that CMB photons undergo along their path from the last scattering surface to us.  This temperature variation arises from the energy shift experienced by photons as they traverse time-varying gravitational potentials. We can retrieve this effect via the integration of the time derivative of $\Phi$ over the whole propagation path~\cite{Giannantonio_2008}:
\begin{equation}
    \frac{\Delta T_{\rm ISWRS}}{T} (\hat{\boldsymbol{n}}) = -2 \, \int {\rm d}z \, \frac{{\rm d}\Phi(\hat{\boldsymbol{n}}, z)}{{\rm d}z} \,,
\label{eq:iswrs}
\end{equation}
where $\hat{\boldsymbol{n}}$ is a unit direction vector on the sphere, $T$ is CMB temperature, $\Phi$ is the gravitational potential obtained from the Poisson equation, and we assume the visibility function of the photons $e^{- \tau(z)}\approx 1$~\cite{Stolzner_2018}.

As anticipated, together with DE, another main responsible for the production of a not-vanishing time-derivative of the gravitational potential is the presence of massive neutrinos. This is because, after becoming non-relativistic, neutrinos free-stream with large thermal velocities that suppress the growth of density perturbations on scales smaller than the so-called ``free-streaming length''~\cite{Lesgourgues_2008,Lesgourgues_2012}:
\begin{equation}
    \lambda_{\rm FS}(z, m_{\nu}) \simeq 8.1 \frac{H_{0}(1+z)}{H(z)} \frac{1\text{ eV}}{ m_{\nu}}\, h^{-1}\text{Mpc} \,,
\end{equation}
where $m_{\nu}$ is the mass of the single neutrino species (i.e. $\nu_{e}, \nu_{\mu} \text{ or } \nu_{\tau}$), $H(z)$ is the Hubble parameter as a function of the redshift $z$, that gives a measure of the expansion rate of the Universe, and $H_{0} \equiv H(z=0)$ is the Hubble constant. Moreover, because of the gravitational back-reaction effects, the evolution of cold dark matter (CDM) and baryon densities is affected as well by the presence of neutrinos, and the total matter power spectrum is suppressed at scales $\lambda \ll \lambda_{\rm FS}$~\cite{Rossi_2014}. Consequently, on small cosmological scales, the free-streaming of neutrinos induces a slow decay of the gravitational potential, acting during both the matter and the DE dominated eras. This effect depends on the total neutrino mass $M_{\nu}$.  The more neutrinos are massive, the more the matter power spectrum will be suppressed with respect to the massless neutrino case, and therefore nonlinearities will appear on smaller scales. 

In the angular power spectrum of the ISWRS effect, this suppression translates into a decrease with respect to the massless scenario, and this decrease intensifies as the neutrino mass increases~\cite{Carbone_2016, Cuozzo_2024}.

\subsection{Cross-correlation with matter tracers in the presence of massive neutrinos}
\label{sec:cross-correlations}
Despite its significant potential to uncover a multitude of insights, the ISWRS effect remains challenging to measure because it is extremely faint compared to CMB primary anisotropies. However, Crittenden and Turok (1996)~\cite{Crittenden_1996} showed that it could be detected through the cross-correlation of the CMB with a local tracer of the mass. 
This is because LSS probes trace the same gravitational potential whose variation induces the ISWRS effect~\cite{Kneissl_1997, Boughn_1998, Boughn_2001, Boughn_2004, Giannantonio_2008, Lesgourgues_2008, Douspis_2008}.
In this work, we focus on galaxy clustering (GC), cosmic shear (CS), and CMB-lensing potential (CMBL). 

\begin{figure*}[!ht]
\centering
\includegraphics[width=0.99\textwidth]{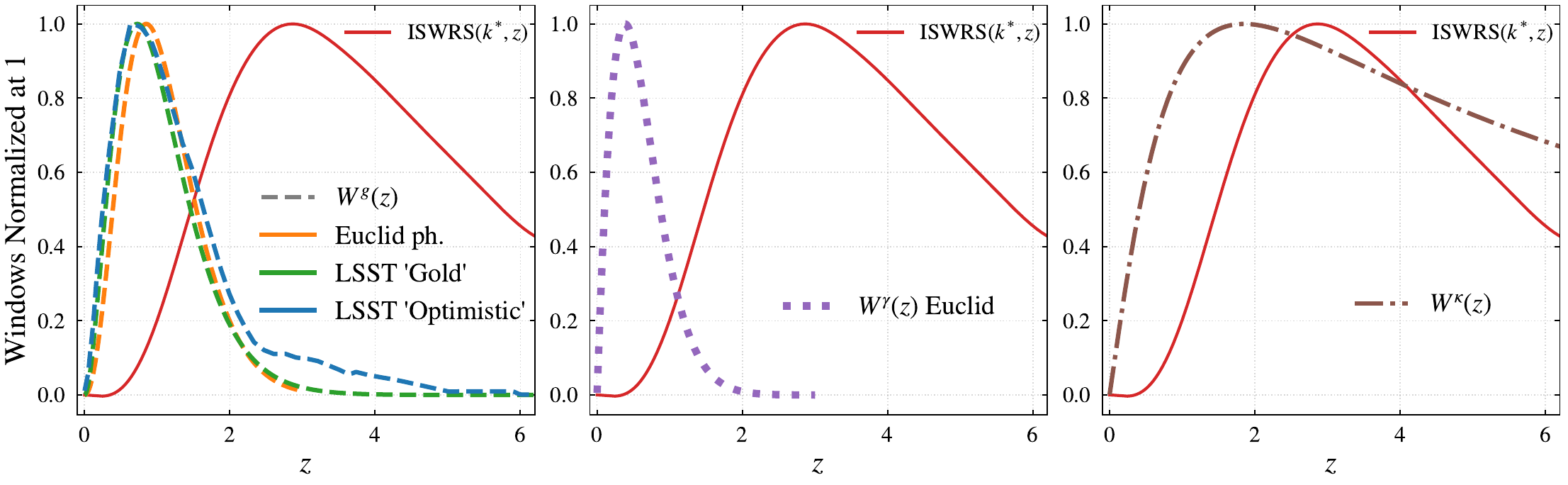}
\caption{Time evolution functions of the investigated fields. The solid red lines in the three panels represent the ISWRS response as a function of $z$ computed at a scale $k^{*} = (\ell^{*} + 1/2)/\chi(z)$, where $\ell^{*} = 5000$. (This specific scale has been choosen since it represents the range of nonlinear scales that mainly contribute to the cumulative $S/N$.) The left panel shows the window function of the Euclid photometric GC sample (dashed orange line), of the LSST  ``Gold'' (dashed green line) and the  LSST ``Optimistic'' GC samples (dashed blue line). The central panel represents the CS window function of the Euclid photometric survey (dotted purple line). Finally, the CMB lensing convergence window function is represented in the right panel (dot-dashed brown line).}  
\label{fig:windows}
\end{figure*}
 Reconstructing the ISWRS effect via its cross-correlation with these probes is possible because they live in the same Universe era in which the ISWRS effect is generated. This means that the window functions of these fields, which represent their evolution in time, overlap, as shown in Figure~\ref{fig:windows}.  The LSS probes window functions can be computed as~\cite{EC20, Piccirilli_23, Smith_2009, PlanckXIII_2016, Planck2018_lensing}:
\begin{subequations}
\label{eq:windows}
\begin{align}
        &W^{g}(z) = \frac{n(z)}{\Bar{n}} \, b(z) \label{eq:win_gc}\,,
        \\
        &W^{\gamma}(z) = \frac{3}{2}  \, \frac{H_{0}^{2}}{c \, H(z)}  \Omega_{m}  \, \frac{\chi(z)}{a(z)} \, \frac{ 1}{\Bar{n}}\,  \int_{z}^{z_{\rm max}} {\rm d}z'\, n(z') \, \frac{\chi(z') - \chi(z)}{\chi(z')} \label{eq:win_cs} \,, 
        \\
        &W^{\kappa}(z) = \frac{3}{2}  \, \frac{H_{0}^{2}}{c \, H(z)} \Omega_{m} \, \frac{\chi(z)}{a(z)} \,  \frac{\chi(z_{ls}) - \chi(z)}{\chi(z_{ls})} \label{eq:win_cl}\,,
        \,
\end{align}
\end{subequations}
where $n(z)$ is the galaxy number density as a function of redshift; $b(z)$ is the galaxy bias function; $a(z)$ is the Universe scale factor; $\Omega_{m}$ is the parameter of matter density today;  $\chi(z)$ is the comoving distance;  $z_{ls} = 1100$ is the redshift at the last-scattering surface and:
\begin{equation*}
    \Bar{n}=\int_{z_{\rm min}}^{z_{\rm max}} {\rm d}z\, n(z) \,.
\end{equation*}
Equations~\eqref{eq:win_gc} and~\eqref{eq:win_cs} represent the window functions for the galaxy distribution ($g$) and shear ($\gamma$), while Equation~\eqref{eq:win_cl} is the convergence ($\kappa$) window function, that can be converted into CMBL potential via Poisson's equation in harmonic space:
\begin{equation}
    \kappa_{\ell m} = \frac{\ell \,(\ell+1)}{2} \, \Phi_{\ell m}\,
    \label{eq:poisson_harmonic}.
\end{equation}
The time evolution of the ISWRS can be instead directly retrieved from the derivative of the Poisson's equation~\cite{Cai_2010}:
\begin{equation}
\dot\Phi(k,t) = - \frac{3}{2} \, \Omega_{m} \, \left( \frac{H_{0}}{c\,k}\right)^{2} \, \partial_{t}\left[\frac{\delta(k,t)}{a(t)}\right]\,.
\label{eq:win_iswrs}
\end{equation}
We explicitly assume a flat Universe. The larger the overlap between the probes time evolution function, the higher their cross-spectrum is expected to be, and consequently the probability of detecting the ISWRS signal is increased. This is because the cross-spectrum depends on the LSS windows product with the ISWRS as follows:
\begin{equation}
    \label{eq:cross-spectrum}
C^{\dot\Phi \text{Y}}_{\ell} =  -2 \int {\rm d}z\, \frac{H(z)}{c\, \chi^{2}(z)} \, W_{\text{Y}}(z) P_{\dot\Phi,\delta}(k,z)  \,,
\end{equation} 
 where the $P_{\delta\dot\Phi}(k,z)$ power spectrum arises after the window functions are multiplied by the $\delta(k,z)$ function and the relation $\langle \dot\Phi(k,z), \delta(k,z)\rangle \equiv P_{\dot\Phi\delta}(k,z)$ is used.
The Y field is one of the $\Phi$-tracer probes ($g$, $\gamma$, $\Phi$), and we get respectively~\cite{Mangilli_2009, Cuozzo_2024, Lesgourgues_2008, Smith_2009, Ferraro_2022}:
\begin{subequations}
\label{eq:cross_cls}
\begin{align}
   C_{\ell}^{\dot\Phi g} &= \frac{3\Omega_{m}H_{0}^{2}}{2c^{3} \, (\ell + \nicefrac{1}{2})^{2}}  \int_{z_{\rm min}}^{z_{\rm max}} {\rm d}z\, H(z) \frac{n(z)}{\Bar{n}} \, b(z) \, a(z) \, \left[\partial_{z}\frac{P_{\delta\delta}(k = \frac{\ell+\nicefrac{1}{2}}{\chi(z)},z)}{a^{2}(z)}\right] \label{eq:tg} \,, \\
    C_{\ell}^{\dot\Phi \gamma} &= \left[\frac{3 \Omega_{m} H_{0}^{2}}{2\, (\ell + \nicefrac{1}{2}) \, c^{2}}\right]^{2} \, \int_{z_{\rm min}}^{z_{\rm max}} {\rm d}z\,\left[\partial_{z}\frac{P_{\delta\delta}(k = \frac{\ell+\nicefrac{1}{2}}{\chi(z)},z)}{a^{2}(z)}\right] \, \times \notag \\ 
    &\hspace{6cm}\times \frac{\chi(z)}{\Bar{n}} \, \int_{z}^{z_{\rm max}} {\rm d}z'\, n(z') \, \frac{\chi(z') - \chi(z)}{\chi(z')} \label{eq:tw} \,, \\
    C_{\ell}^{\dot\Phi\Phi} &= 2 \left[\frac{3 \Omega_{m} H_{0}^{2}}{2\, (\ell + \nicefrac{1}{2})^{2} \, c^{2}}\right]^{2} \int_{z_{\rm min}}^{z_{\rm max}} {\rm d}z\, \chi(z) \,\frac{\chi(z_{ls}) - \chi(z)}{\chi(z_{ls})} \left[\partial_{z} \frac{P_{\delta\delta}(k = \frac{\ell+\nicefrac{1}{2}}{\chi(z)},z)}{a^{2}(z)}\right] \label{eq:tp} \,,
\end{align}
\end{subequations}
where $k$ is in Limber approximation~\cite{Limber_53}:
\begin{equation}
    k_{Limber}=\frac{\ell+1/2}{\chi(z)}\,,
\end{equation}
and the derivation of $\partial_{z}P_{\delta\delta}(k,z)$ is detailed in Appendix~\ref{app:ccs}. Moreover, in Equation~\eqref{eq:tp}, the following approximation has been used for Equation~\eqref{eq:poisson_harmonic}:
\begin{equation}
    \Phi_{\ell m} \simeq \frac{2}{(\ell + \nicefrac{1}{2})^2} \, \kappa_{\ell m}\,,
    \label{eq:poisson_harmonic_prox}
\end{equation}
valid since we are focussing on very small scales (i.e. very large $\ell$) in this work.

\begin{figure*}[!ht]
\centering
\includegraphics[width=0.99\textwidth]{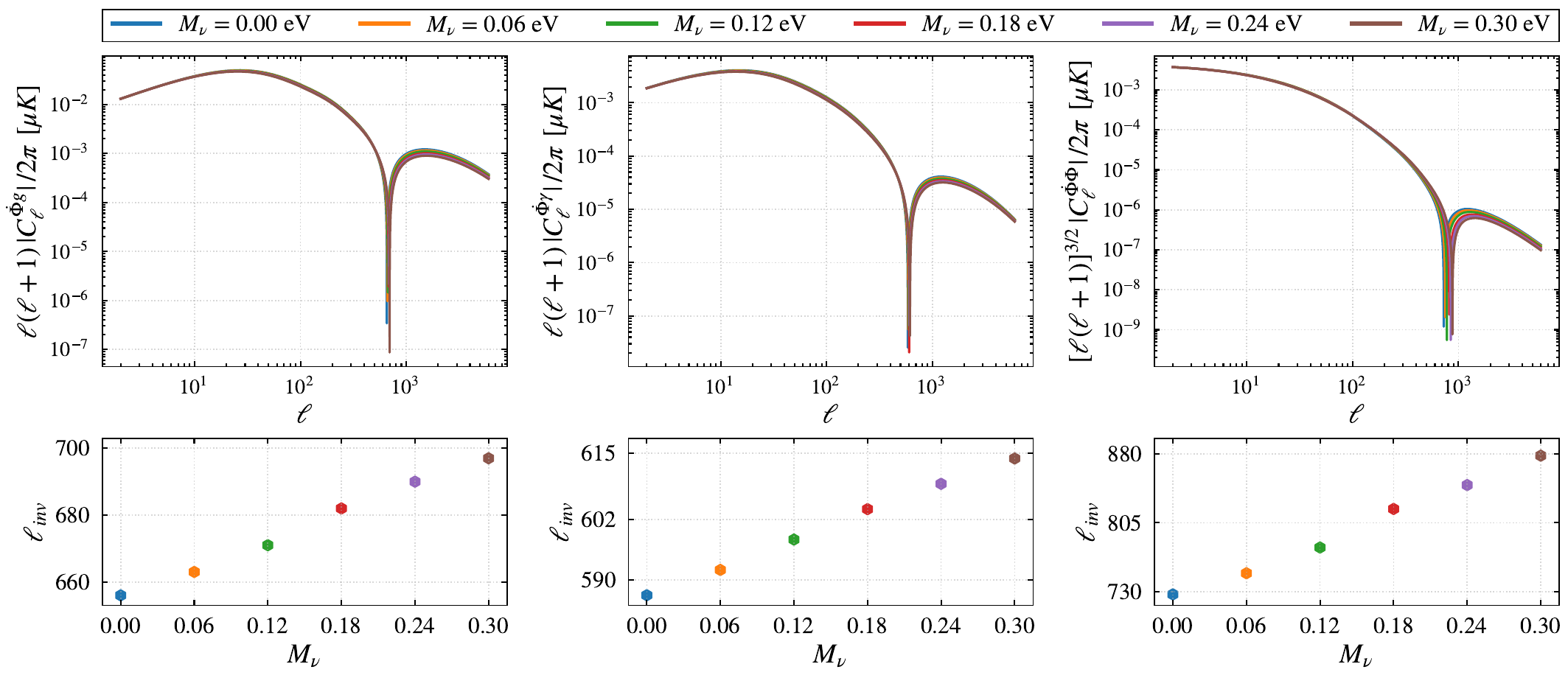}
\caption{Analytical predictions for the cross-correlation between the ISWRS effect and GC, CS and CMBL, from left to right, computed for $\nu\Lambda$CDM models with five different values of the total neutrino mass: $M_{\nu} = 0.06 \text{ eV (orange)}, 0.12\text{ eV (green)}, 0.18\text{ eV (red)}, 0.24\text{ eV (purple)}, 0.30\text{ eV  (brown)}$. Together with them, the results for the massless case have been reported (blue). As GC and CS probes, the Euclid photometric sample has been used~\cite{EC20}.
The upper panels represent the absolute value of the three angular cross-power spectra. The lower ones show how the position of the characteristic sign inversion varies with the value of $M_{\nu}$.}
\label{fig:cross-spectra}
\end{figure*}

The curves of Figure~\ref{fig:cross-spectra} represent the absolute value of Equations~\eqref{eq:cross_cls}, from left to right. To realise them, and all the results shown in this work, we use as a baseline the Planck 2018 cosmology (parameters from the last column of Table 2 of~\cite{PlanckVI_2020}), namely a flat $\Lambda$CDM model with $M_{\nu} = 0.06$ eV generalised by varying only the sum of the neutrino masses over the values $M_{\nu} = [0.0, 0.06, 0.12, 0.18, 0.24, 0.30]$ eV, whence the corresponding values of $\Omega_{\nu}$ and $\Omega_{cdm}$, keeping fixed $\Omega_{m}$ and $\Omega_{b}$. The galaxy selection and bias functions are those of the Euclid photometric survey, given respectively by~\cite{EC20, Tutusaus_2020}:
\begin{subequations}
\begin{align}
    n_{\text{Euclid}}(z) &\propto \left(\frac{z}{z_{0}}\right)^{2} \text{exp}\left[-\left(\frac{z}{z_{0}}\right)^{3/2} \right] \label{eq:euclid_n} \,,\\
    b_{\text{Euclid}}(z) &=   A + \frac{B}{1 + \exp[-(z-D){C}]} \label{eq:euclid_b} \,,
\end{align}
\end{subequations}
with $z_{0} = 0.9/\sqrt{2}$, $A=1.0$, $B=2.5$, $C=2.8$, $D=1.6$, and mean galaxy number density of $\Bar{n} = 30\text{ arcmin}^{-2}$. 

Additionally, in this work, we test the galaxy selection and bias functions of the LSST survey~\cite{LSST_2009,LSST_2018} (see left panel of Figure~\ref{fig:windows}):
\begin{subequations}
\begin{align}
    n_{\text{LSST}}(z) &\propto \frac{1}{2 z_{0}}\left(\frac{z}{z_{0}}\right)^{2} \text{exp}\left[-\left(\frac{z}{z_{0}}\right) \right] \label{eq:lsst_n}\,,\\
    b_{\text{LSST}}(z) &= 0.95 \, (1+z)\label{eq:lsst_b}\,,
\end{align}
\end{subequations}
where $z_{0} = 0.3$. These functions refer to the LSST ``Gold'' sample, where with the term ``Gold'' we consider the $i< 25$ magnitude sample, with $\Bar{n} = 40\text{ arcmin}^{-2}$~\cite{LSST_2009}. We consider even a more ``Optimistic'' LSST sample corresponding to $i< 27$~\cite{Gorecki_2014}, which includes Lyman break galaxies from redshift dropouts, which results in an increase of the number density of galaxies to $\Bar{n} = 66\text{ arcmin}^{-2}$~\cite{Yu_2018, Yu_2022,Ferraro_2022}.

All the other terms in Equations~\eqref{eq:cross_cls} have been computed using \texttt{CAMB}\footnote{\url{https://CAMB.readthedocs.io/en/latest/}}~\cite{Lewis_2011}.

The cross-spectra are shown in Figure~\ref{fig:cross-spectra} and are characterised by the presence of a sign inversion that appears in correspondence with the transition from the linear to the nonlinear regime, due to the anti-correlation between RS and the $\Phi$ tracers (i.e.  $\langle \Phi\dot{\Phi}\rangle > 0$ in the late-time ISW regime, and $\langle \Phi\dot{\Phi}\rangle < 0$ in the RS regime ~\cite{Nishizawa_2014, Nishizawa_2008}). Moreover, the presence of massive neutrinos affects these cross-angular power spectra. In fact, as neutrino mass increases, nonlinearities appear on smaller cosmological scales and the sign inversion appears on larger multipoles~\cite{Carbone_2016,Cuozzo_2024}.  From the subpanels of Figure~\ref{fig:cross-spectra} it is in fact clear that the more neutrinos are massive, the more the sign inversion shifts towards smaller cosmological scales, and the dependency is almost linear~\cite{Cuozzo_2024}.

For our analyses, in \texttt{CAMB} we have adopted the Takahashi Halofit model~\cite{Smith_2003,Takahashi_2012}, that includes Bird's\footnote{As reported in the Readme of the \texttt{CAMB} webpage, on March 2014 modified massive neutrino parameters were implemented in the nonlinear fitting of the total matter power spectrum to improve the accuracy of the updated \texttt{Halofit} version from~\cite{Takahashi_2012}. These fitting parameters, accounting for nonlinear corrections in the presence of massive neutrinos, are different from the ones
implemented by~\cite{Bird_2012} in the original Halofit version from \cite{Smith_2003}.} corrections for neutrinos masses~\cite{Bird_2012} (\citetalias{Takahashi_2012} from now on). The choice to use this nonlinear model follows from our previous work~\cite{Cuozzo_2024}, where we validated Equations~\eqref{eq:tg} and \eqref{eq:tp} against the cross-spectra extracted from the ``Dark Energy and Massive Neutrino Universe'' (DEMNUni) N-body numerical simulations~\cite{Carbone_2016, Castorina_2015,Moresco_2017,Zennaro_2018,Ruggeri_2018,Bel_2019,Parimbelli_2021,Parimbelli_2022, Baratta_2022, Guidi_2022, Gouyou_Beauchamps_2023, Carella_in_prep}, and verified that the \citetalias{Takahashi_2012} modelling works better for these analytical reconstructions if compared with the other model implemented in \texttt{CAMB}, the Mead2020 model~\cite{Mead_2021} (\citetalias{Mead_2021} from now on). Precisely, in~\cite{Cuozzo_2024} we assessed how non-linear Halofit models affect the determination of the sign-inversion multipole ($\ell_{inv}$) in the analyzed cross-correlations, which is the key feature constraining the assumed $M_{\nu}$. There we verified that when using \citetalias{Takahashi_2012}, for small values of the neutrino masses, we find an agreement with the DEMNUni simulations at $\lsim 1\sigma$ level for the cross-correlation between the ISWRS and the CMBL and at $\sim 1-1.5\sigma$ level for the cross-correlation with the galaxy distribution.  Conversely, the \citetalias{Mead_2021} model shows differences at $\gsim 2\sigma$ level for both cross-correlations.
For the validation of Equation~\eqref{eq:tw}, see Appendix~\ref{app:validation}.

\section{Detectability of the ISWRS effect and identification of \texorpdfstring{$\nu\Lambda$}CDM model}
\label{sec:results}
As previously mentioned, properly reconstructing the ISWRS effect via its cross-correlation with GC, CS and CMBL is possible because these probes trace the variation of the gravitational potential that induces the ISWRS itself. Since the amplitude of the LSS and ISWRS cross-angular power spectra directly depends on the overlap between their window functions, the larger the overlap, the greater their cross-spectrum is expected to be. The straightforward consequence is an increased probability of detecting the ISWRS signal.

The significance of the detection can be quantified by the cumulative signal-to-noise ratio $(S/N)$, where the signal is the cross-power spectrum $C_{\ell}^{\dot\Phi Y}$ (with $Y$ one of the $\Phi$ tracers: $g$, $\gamma$, $\Phi$), and the associated uncertainty is given by:
\begin{equation}
    \sigma_{\ell}^{2} =
    \frac{(C_{\ell}^{TT}+d_{\ell})(C_{\ell}^{YY}+N^{Y}_{\ell})+ (C_{\ell}^{\dot\Phi Y})^{2}}{ (2\ell + 1) \,  f_{sky}}\,,
\label{eq:covariance}
\end{equation}
resulting in~\citep{Stolzner_2018, Ferraro_2022}:
\begin{equation}
    {\rm CUM.}\left(\frac{S}{N}\right)(\ell_{\text{max}}) \, =\, \sqrt{ f_{sky} \, \sum_{\ell=\ell_{\text{min}}}^{\ell_{\text{max}}}
    \frac{ (2\ell + 1) \, (C_{\ell}^{\dot\Phi Y})^{2}}{(C_{\ell}^{TT}+d_{\ell})(C_{\ell}^{YY}+N^{Y}_{\ell})+ (C_{\ell}^{\dot\Phi Y})^{2}}} \,,
\label{eq:snr}
\end{equation}
where $f_{sky}$ is the sky fraction covered by the overlap of the CMB experiment and LSS survey, $C_{\ell}^{TT}$ is the lensed primary CMB power spectrum, $d_{\ell}$ is the CMB detector noise (to which foreground residuals can be added), $N^{Y}_{\ell}$ is the noise associated to the Y probe. 

Because of the different impact that different values of $M_{\nu}$ have at very small scales on the cross-spectra between the ISWRS effect and the three analysed probes (shown in the lower panel of Figure~\ref{fig:cross-spectra}), a high significance detection for these correlations can in principle even allow to disentangle different $\nu\Lambda$CDM models. We can, in fact, quantify the detectability, in units of $\sigma$, of a signal where $M_{\nu}>0$ with respect to the massless case, via:
\begin{equation}
     \sqrt{\chi^2}(\ell_{\text{max}}) = \sqrt{\sum_{\ell=\ell_{\text{min}}}^{\ell_{\text{max}}} \frac{[C^{\dot\Phi\text{Y}}_{\ell}(M_{\nu} = 0.0\text{ eV}) - C^{\dot\Phi\text{Y}}_{\ell}(M_{\nu} > 0.0\text{ eV})]^{2}}{\sigma^{2}_{\ell}(M_{\nu} = 0.0\text{ eV})}} \,.
\label{eq:mnu}
\end{equation}

In this work, $\ell_{\text{min}} = 2$, $\ell_{\text{max}} = 6000$~\citep{Ferraro_2022}, $C_{\ell}^{\dot\Phi Y}$ has been computed with Equations~\eqref{eq:cross_cls}, $C_{\ell}^{TT}$ has been extracted from \texttt{CAMB} and $C_{\ell}^{YY}$ has been computed analytically as~\citep{Piccirilli_23}:
\begin{equation}
C^{YY}_{\ell} =  \int {\rm d}z\, \frac{H(z)}{c\, \chi^{2}(z)} \, [W^{Y}(k,z)]^{2} \, P_{\delta\delta}(k,z)  \,.
\label{eq:auto_cl}
\end{equation}
which is Equation~\eqref{eq:cross-spectrum} in the case of $X=Y$.

\subsection{Ideal forecast}
\label{sec:7_idealcase}
Despite the significant overlap between the window functions shown in Figure~\ref{fig:windows} which makes these cross-correlation signals particularly robust, the current CMB and LSS surveys are characterised by levels of statistical and instrumental noises that, together with the presence of foreground contamination, make the detectability of the ISWRS challenging, and the possibility of disentangling between different $\nu\Lambda$CDM models likely unfeasible~\cite{Cuozzo_2024}. 

Ideally, the best detection (i.e., the highest $S/N$) would be possible if the $Y$ probe window function perfectly overlaps with the ISWRS one, so that $C_{\ell}^{\dot\Phi Y} \equiv C_{\ell}^{\dot\Phi\dot\Phi}$, the survey is full-sky ($f_{sky} \simeq 1$), and the statistical and instrumental noises are negligible. 

In such a case, Equation~\eqref{eq:snr} becomes~\citep{Ferraro_2022}:
\begin{equation}    
     {\rm CUM.}\left(\frac{S}{N}\right)_{\text{ideal}}(\ell_{\text{max}}) = \sqrt{\sum_{\ell=\ell_{\text{min}}}^{\ell_{\text{max}}}
     \frac{(2\ell+1) \, (C_{\ell}^{\dot\Phi\dot\Phi})^{2}}{(C_{\ell}^{TT} \cdot C_{\ell}^{\dot\Phi\dot\Phi})+ (C_{\ell}^{\dot\Phi\dot\Phi})^{2}}} \,.
\label{eq:snr_ideal}
\end{equation}
Differently from the other auto-power spectra used in this work, $C_{\ell}^{\dot\Phi\dot\Phi}$ cannot be computed using Equation~\eqref{eq:auto_cl} without a proper treatment of nonlinearities.  This is because the current status of Halofit models~\cite{Smith_2003} in Boltzmann solver codes like \texttt{CAMB} is not advanced enough to accurately reconstruct $P_{\dot\Phi\dot\Phi}(k,z)$ (see Appendix~\ref{app:ccs}) in the nonlinear regime. To properly compute the ISWRS spectrum, including corrections due to nonlinearly evolved potentials, one needs to perform an expansion of the dark matter power spectrum $P_{\delta\delta}(k,z)$ using perturbation theory methods~\cite{Shafer_2011, Smith_2009}, or to fit the $P_{\delta\delta}(k,z)$ model to N-body simulations~\cite{Cai_2009}, or to directly use the result of N-body simulations, as we show in this work.

Here we use the ISWRS auto-power spectra extracted from the DEMNUni maps~\citep{Carbone_2016} realised in the redshift range $z=[0.02,7]$ for the baseline Planck Collaboration (2014b) cosmology~\cite{PlanckXVI_2014}, generalised to three $\nu\Lambda$CDM cosmologies with $M_{\nu}=0.0, 0.17, 0.30$ eV.
\begin{figure*}[!ht]
\centering
\includegraphics[width=0.7\textwidth]{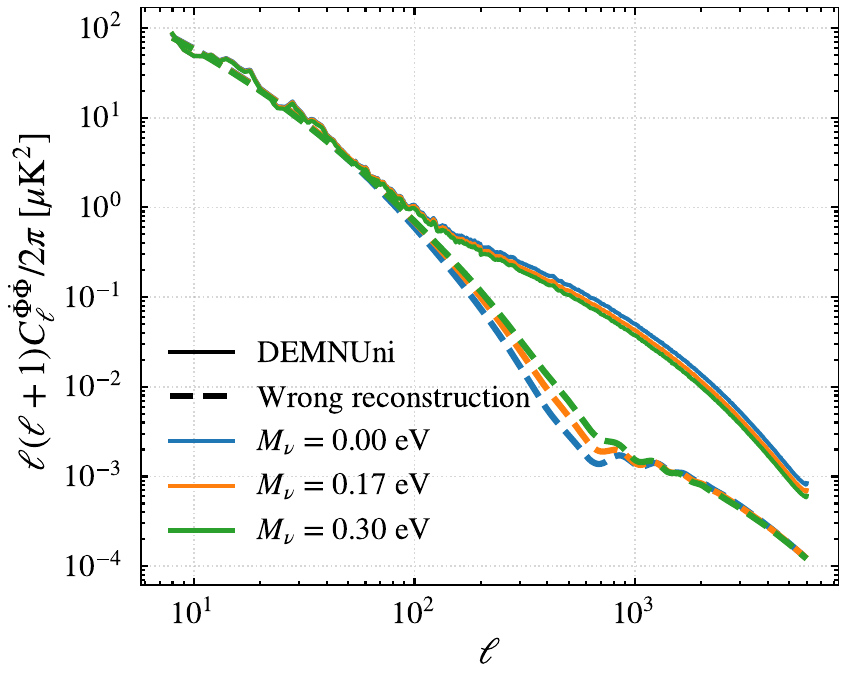}
\caption{Comparison between the ISWRS angular auto-spectra extracted from DEMNUni simulations in the range $z=[0.02, 7]$ (solid lines) and those wrongly analytically computed (dashed lines), for three $M_{\nu}$ values: $0.0$ eV (blue),  $0.17$ eV (orange),  $0.30$ eV (green).}
\label{fig:iswrs_spectra}
\end{figure*}

As shown in Figure~\ref{fig:iswrs_spectra}, when using Equation~\eqref{eq:auto_cl} with $P_{\delta\delta}(k,z)$ computed with \texttt{CAMB}, as done in previous work ~\cite{Ferraro_2022}, we are unable to properly recover nonlinear scales, underestimating the signal by more than 2 orders of magnitude.

In what follows, we therefore use the $C_{\ell}^{\dot\Phi\dot\Phi}$ extracted from DEMNUni maps in Equation~\eqref{eq:snr_ideal} to estimate the maximum achievable S/N ratio. We consider both the cases of lensed and delensed $C_{\ell}^{TT}$~\cite{Hotinli_2022}, and assess the possibility of disentangling different values of $M_{\nu}$ with Equation~\eqref{eq:mnu}.

 \begin{figure*}[!ht]
\centering
\includegraphics[width=0.95\textwidth]{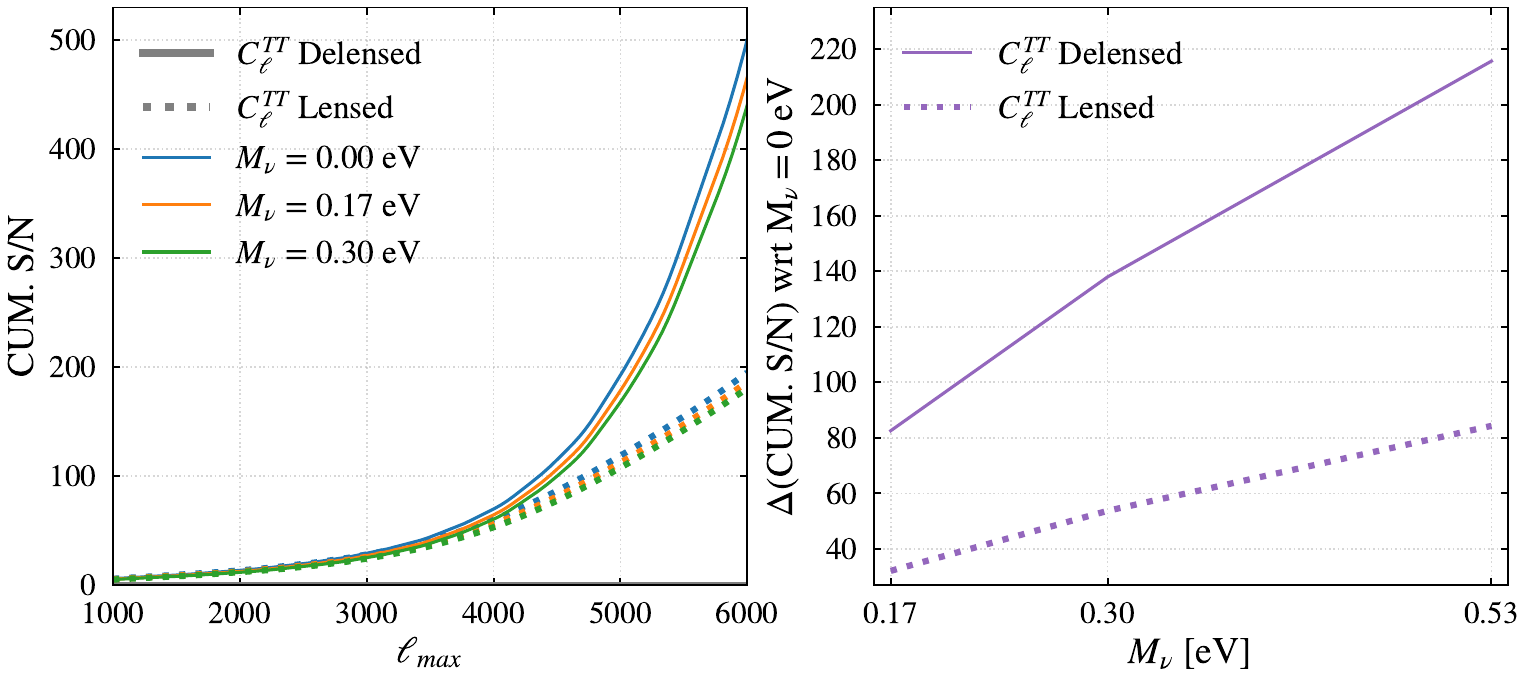}
\caption{The left panel represents the results of the cumulative signal-to-noise ratio (CUM. $S/N$) obtained using Equation~\eqref{eq:snr_ideal} with the $C_{\ell}^{\dot\Phi\dot\Phi}$ extracted from DEMNUni maps with $M_{\nu}$ = $0.0$ eV (blue), $0.17$  eV (orange), $0.30$ eV (green). The right panel shows the results for the identification of the $\nu\Lambda$CDM model with respect to the massless case, obtained using Equation~\eqref{eq:mnu}. We also add the result for the $\nu\Lambda$CDM cosmology with $M_{\nu}=0.53$ eV to show the actual increasing trend as the neutrino mass increases. Both panels present the results obtained with  lensed (solid lines) and delensed (dotted lines) $C_{\ell}^{TT}$.}
\label{fig:ideal_case}
\end{figure*}

As shown in the left panel of Figure~\ref{fig:ideal_case},  a signal-to-noise $\sim 200$ is found, which can increase up to $\gsim 500$ if delensing techniques are applied to the CMB anisotropy data. In these conditions, it would be possible to identify the $\nu\Lambda$CDM model. The ability to distinguish it from the massless case ranges from approximately $30\sigma$ for $M_{\nu} = 0.17$ eV without delensing, to about $140\sigma$ for $M_{\nu} = 0.30$ eV with delensed CMB, as shown in the right panel of Figure~\ref{fig:ideal_case}. To emphasise that the trend does not follow a specific linear growth law, in the right panel of Figure~\ref{fig:ideal_case} we even show the result for $M_{\nu}=0.53$ eV, made possible by the availability of the DEMNUni suite.

\subsection{Optimistic forecast}
Although the ideal case is certainly promising, it remains inapplicable to our Universe. Consequently, we have chosen to introduce the analysis of an optimistic yet physically motivated benchmark case.

Specifically, we reconstruct GC, CS and CMBL cross-spectra with ISWRS for three $\nu\Lambda$CDM cosmologies with $M_{\nu}=0.06, 0.18, 0.30$ eV and estimate their cumulative $S/N$ using Equation~\eqref{eq:snr}, with both lensed and delensed $C_{\ell}^{TT}$, assuming full sky coverage $f_{sky} = 1$, no detector noise nor foreground residuals in CMB data $d_{\ell} = 0$, and adopting an optimistic galaxy abundance scenario based on \cite{Conselice_2016}. 

\subsubsection{GC $\times$ ISWRS}
\begin{figure*}[!ht]
\centering
\hspace{-0.83cm}
\begin{tabular}{cc}
\includegraphics[width=0.49\textwidth]{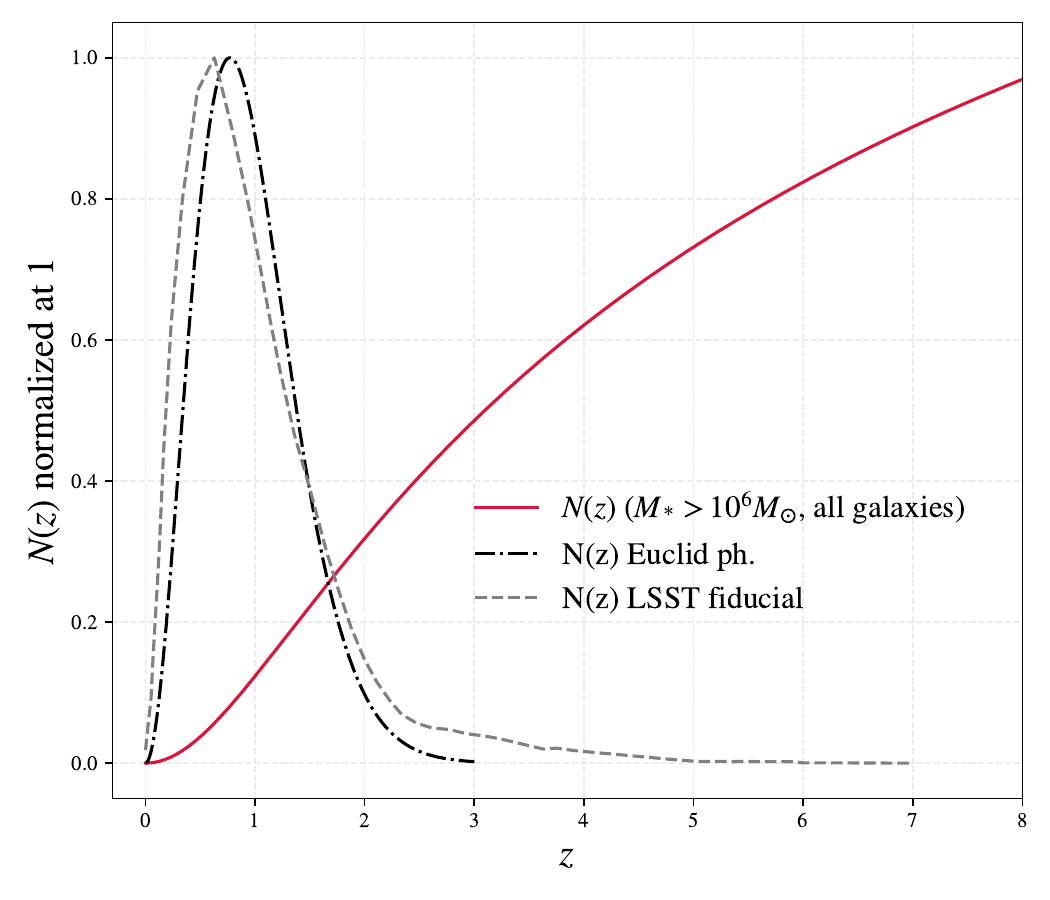} &
    \includegraphics[width=0.503\textwidth]{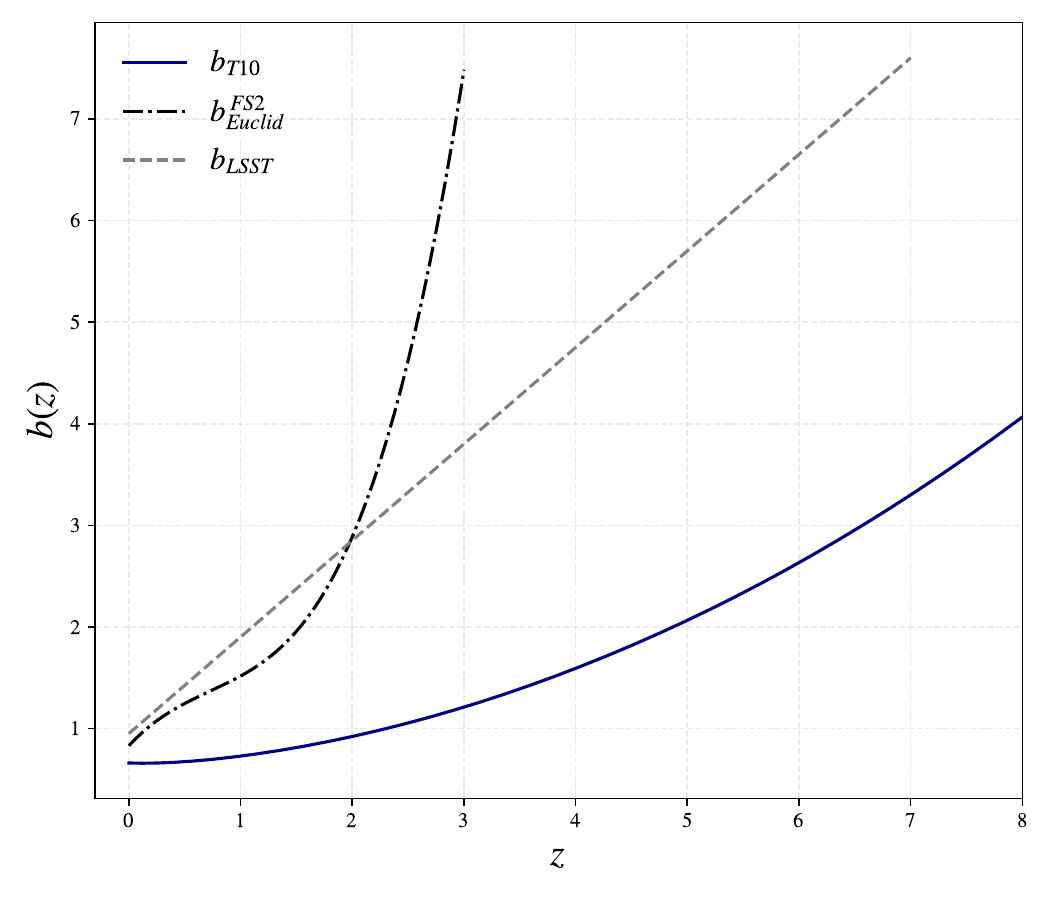}
\end{tabular}
\caption{The left panel shows the galaxy number density divided by their maximum retrieved from \cite{Conselice_2016} (red solid line) compared with the Euclid one (black dot-dashed line), and the LSST fiducial one (grey dashed line). The right panel shows the effective bias obtained from the numerical calibration of~\cite{Tinker_2010} (T10) and computed assuming a minimum halo mass threshold of $M_{\rm h, min} \approx 10^9 \, h^{-1} M_{\odot}$ (navy solid line), compared with Euclid bias (black dot-dashed line) and the LSST one (grey dashed line).}
\label{fig:new_nbz}
\end{figure*}

To maximize the statistical significance of this cross-correlation at high redshifts, where the RS effect is strongest, we adopt the integrated galaxy number density $N(z)$ derived from the comoving density framework of \cite{Conselice_2016}, where the total comoving number density $\phi_{\rm T}(z)$ is computed by integrating the stellar mass function  $\phi(M_*, z)$ down to a faint-end integration boundary of $M_{*, \rm min} = 10^6 \, M_{\odot}$:
\begin{equation}
\phi_{\rm T}(z) = \int_{M_{*, \rm min}}^{\infty} \phi(M_*, z) \, dM_* = \phi_0 \left( \frac{t_0}{t(z)} \right),
\end{equation}
where $M_*$ denotes the galaxy stellar mass, $\phi_{0}$ is the total comoving number density of galaxies at the present epoch, $t(z)$ the cosmic age at a given epoch and $t_0$ the age of the Universe today. To convert this comoving volumetric density into the total number of observable galaxies per unit redshift interval across the full sky, $N(z)$, we weight $\phi_{\rm T}(z)$ by the differential comoving volume element $dV/dz$ as follows:
\begin{equation}
N(z) = \phi_{\rm T}(z) \frac{dV}{dz} = \phi_{\rm T}(z) \cdot 4\pi \left[ \frac{c \, r^2(z)}{H(z)} \right],
\end{equation}
where $r(z)$ is the comoving distance to redshift $z$, $H(z)$ is the Hubble parameter, and $c$ is the speed of light. The result of our computation is reported in Figure~\ref{fig:new_nbz}.

In this low-mass regime, the comoving number density evolves inversely with cosmic time ($\phi_{\rm T} \propto t^{-1}$), tracking the early evolutionary phase where today's massive structures were highly fragmented into a vast population of primordial proto-galaxies. This choice provides a high, sustained number density of tracers well into the Cosmic Dawn, drastically suppressing the Poissonian shot-noise configurations that traditionally plague high-redshift clustering measurements and obscure the faint RS signal.

Alongside the number density, to trace the clustering of such low-mass proto-galaxies, we estimate the associated effective halo bias by assuming that these objects reside in host dark matter halos above a minimum halo mass threshold of $M_{\rm h, min} \approx 10^9 \, h^{-1} M_{\odot}$. This threshold is physically motivated by the atomic hydrogen cooling limit required to sustain early star formation, and is consistent with the empirical stellar-to-halo mass relations (SMHM) at $z > 4$ \cite{Behroozi_2019}, effectively anchoring the $M_{*, \rm min} \sim 10^6 \, M_{\odot}$ stellar mass boundary to its underlying dark matter potential wells.

Because galaxies populate halos across a broad range of masses, we compute the effective linear bias $b(z)$ by integrating the mass-dependent halo bias $b(M, z)$ over the halo mass function, from the minimum halo mass threshold of $M_{\rm h, min} \approx 10^9 \, h^{-1} M_{\odot}$ up to $M_{\rm h, min} \approx 10^{16} \, h^{-1} M_{\odot}$.

The mass dependent integrand bias is evaluated using the numerical calibration from \cite{Tinker_2010}, parameterized in terms of the linear density peak height $\nu(M, z) \equiv \delta_{\rm c} / \sigma(M, z)$ as:
\begin{equation}
b(\nu) = 1 - A \frac{\nu^a}{\nu^a + \delta_{\rm c}^a} + B \nu^b + C \nu^c,
\label{eq:tinker_bias}
\end{equation}
where $\delta_{\rm c} = 1.686$ is the critical density threshold for spherical collapse and $\sigma(M, z)$ is the root-mean-square linear matter fluctuation smoothed on the Lagrangian scale corresponding to halo mass $M$. For halos identified at the canonical overdensity threshold $\Delta = 200$, the calibrated best-fit parameters are $A = 1.0$, $a = 0.132$, $B = 0.183$, $b = 1.5$, $C = 0.265$, and $c = 2.4$. The result is reported in Figure~\ref{fig:new_nbz}.

We note that recent numerical and analytical studies, such as \cite{Wang_2026}, have pointed out percent-level corrections to standard bias prescriptions calibrated at low-$z$, reporting deviations on the order of $3\text{--}4\%$ for $z \ge 6$. However, given the broad redshift integration range adopted in this work, such percent-level variations remain well within the expected observational uncertainties and do not significantly impact our forecast constraints.

\begin{figure*}[!ht]
\centering
\includegraphics[width=0.95\textwidth]{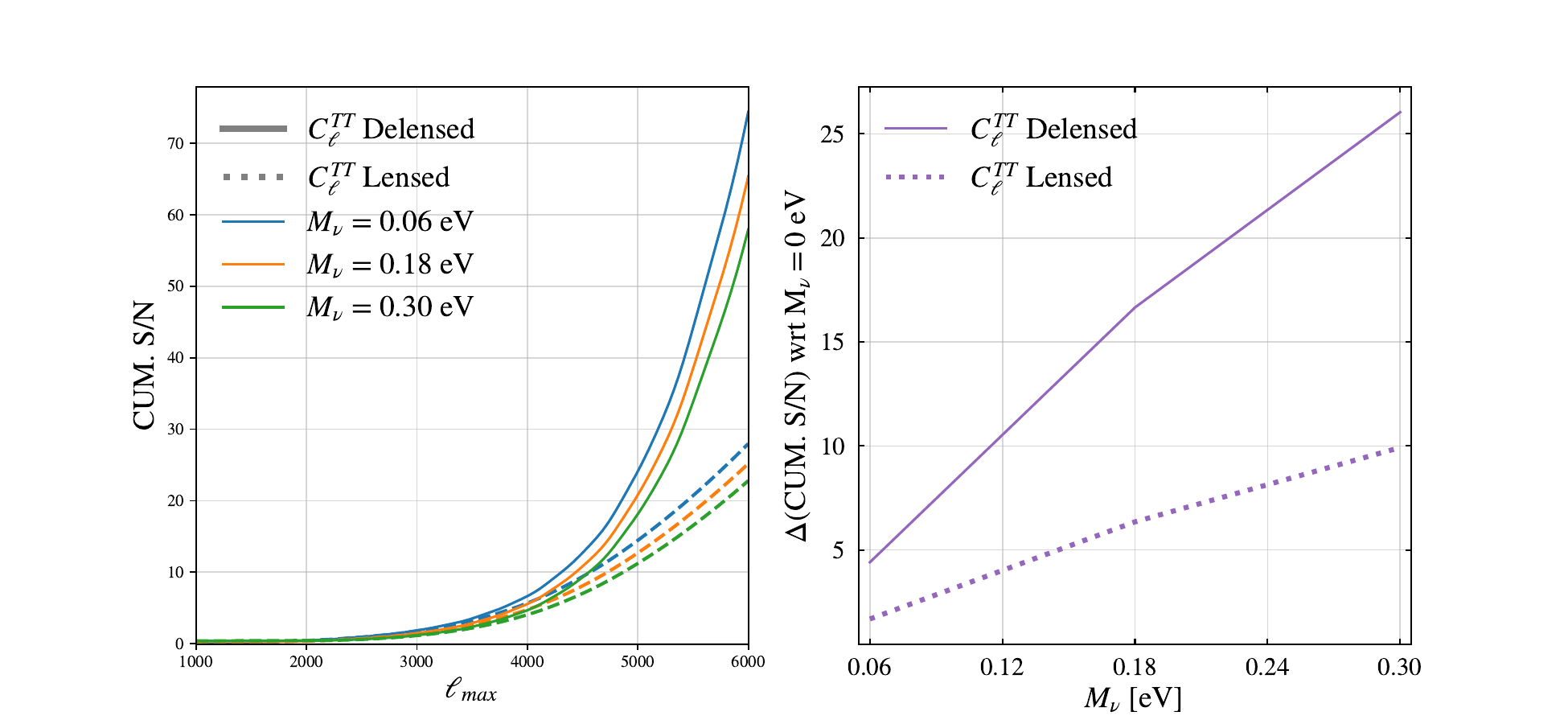}
\caption{The left panel represents the results of the cumulative signal-to-noise ratio (CUM. $S/N$) of $C_{\ell}^{\dot{\Phi}g}$ obtained using Equation~\eqref{eq:snr} with the $C_{\ell}^{gg}$ realised using $N(z)$ from~\cite{Conselice_2016} and $b(z)$ from~\cite{Tinker_2010},  with $M_{\nu}$ = $0.06$ eV (blue), $0.18$  eV (orange), $0.30$ eV (green). The right panel shows the results for the identification of the $\nu\Lambda$CDM model with respect to the massless case, obtained using Equation~\eqref{eq:mnu}. Both panels present the results obtained with  lensed (solid lines) and delensed (dotted lines) $C_{\ell}^{TT}$.}
\label{fig:ideal_gc}
\end{figure*}

Assuming these features for the galaxy distribution,  with a mean galaxy number density of $\Bar{n} = 13.6\times10^{3} \text{ arcmin}^{-2}$~\cite{Conselice_2016}, we compute the cross-correlation spectra between GC and ISWRS and use Equation~\eqref{eq:snr} and Equation~\eqref{eq:mnu} to determine their detectability. The results are reported in Figure~\ref{fig:ideal_gc}.

As shown in the left panel of Figure~\ref{fig:ideal_gc}, the total signal-to-noise ratio ranges from $\sim22$ to $\sim28$, and can increase up to $\sim 75$ when delensing techniques are applied to the CMB anisotropy data. In these conditions, the identification of the $\nu\Lambda$CDM model would be feasible at high significance and the ability to distinguish the model from the massless case reaches few $\sigma$s for $M_{\nu}\gsim0.06$ eV, even when no delensing techniques are applied.

\subsubsection{CS $\times$ ISWRS}
\begin{figure*}[!ht]
\centering
\includegraphics[width=0.95\textwidth]{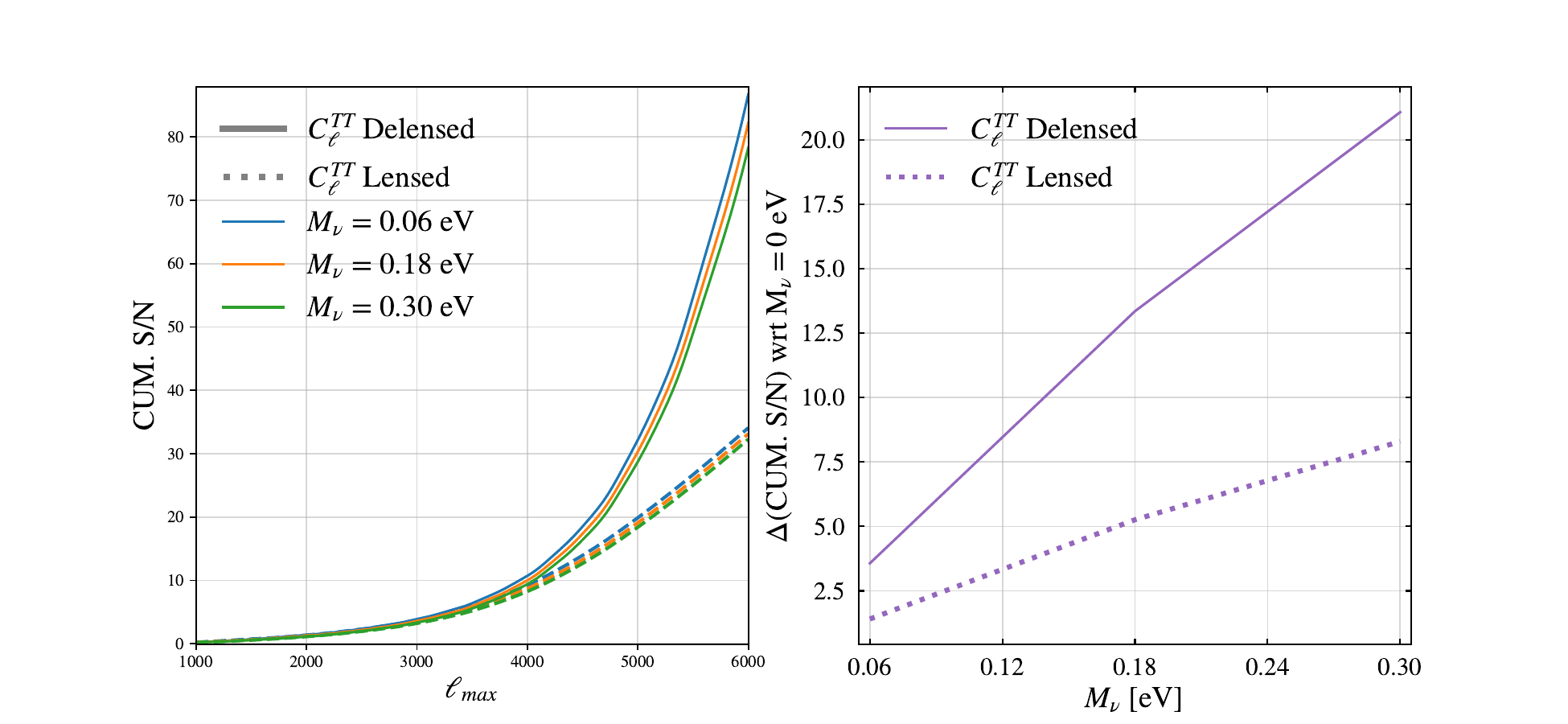}
\caption{Same as Figure~\ref{fig:ideal_gc}, but for $C_{\ell}^{\dot{\Phi}\gamma}$, obtained with the $C_{\ell}^{\gamma\gamma}$ realised using $N(z)$ from~\cite{Conselice_2016}.}
\label{fig:ideal_wl}
\end{figure*}

For this cross-correlation, we reconstruct the weak-lensing spectra using the galaxy number density $N(z)$ from~\cite{Conselice_2016}, introduced in the previous section. The results of the cross-spectra detectability are reported in Figure~\ref{fig:ideal_wl}.

As shown in the left panel of Figure~\ref{fig:ideal_wl}, the total signal-to-noise ratio ranges from $\sim30$ to $\sim35$, but can increase up to $\sim 80$ when delensing techniques are applied to the CMB anisotropy data. In these conditions, the identification of the $\nu\Lambda$CDM model would be feasible at high-significance. Moreover, even in this case, the ability to distinguish the model from the massless case would be possible, even for the most challenging $M_{\nu}=0.06$ eV case.

\begin{figure*}[!ht]
\centering
\includegraphics[width=0.49\textwidth]{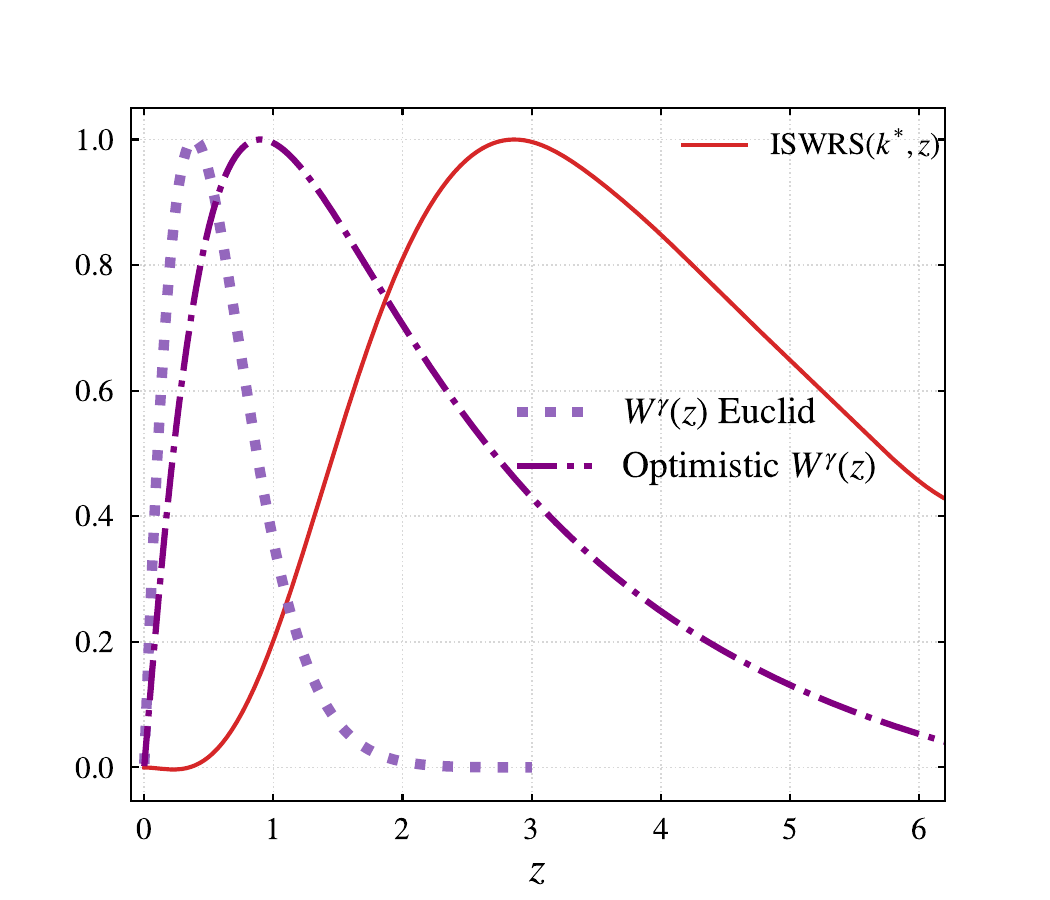}
\caption{Same as the central panel of Figure~\ref{fig:windows}, but with the addition of the optimistic CS window function obtained using $N(z)$ from~\cite{Conselice_2016} (dot-dashed purple line).}
\label{fig:wl_new}
\end{figure*}

This cross-correlation under optimistic conditions, yields impressive results, especially when compared to the realistic results for the same cross-correlation that will be presented in Section~\ref{sec:cross-cs}. This is due to the greater overlap between the ISWRS and the CS window function obtained with $N(z)$ from~\cite{Conselice_2016}, as shown in Figure~\ref{fig:wl_new}, as well as to the significantly higher galaxy number density in this optimistic case. 

\subsubsection{CMBL $\times$ ISWRS}
\begin{figure*}[!ht]
\centering
\includegraphics[width=0.95\textwidth]{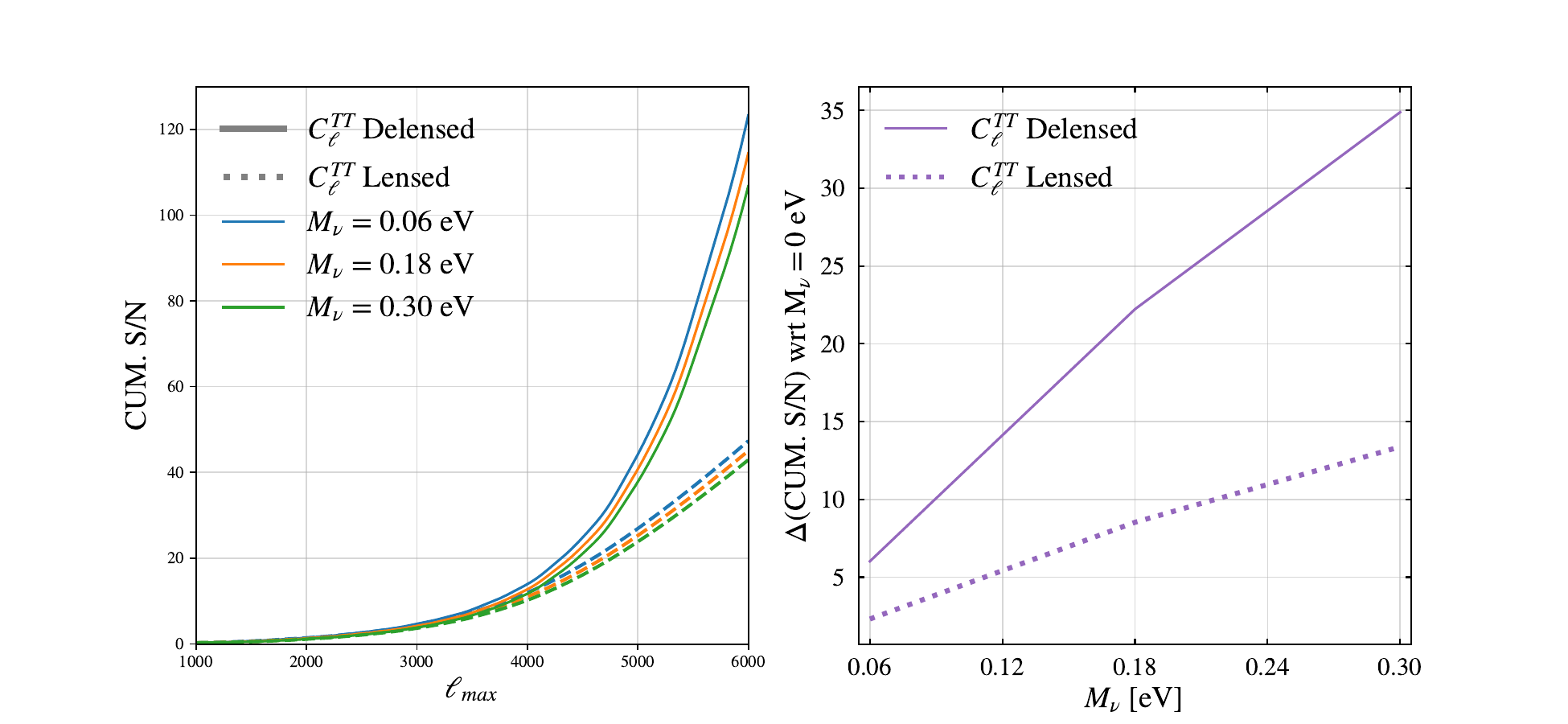}
\caption{Same as Figure~\ref{fig:ideal_gc}, but realised with $C_{\ell}^{\dot{\Phi}\Phi}$.}
\label{fig:ideal_cmbl}
\end{figure*}

For this cross-correlation, in the optimistic scenario, we specifically ignore the CMB lensing reconstruction noise. As shown in the left panel of Figure~\ref{fig:ideal_cmbl}, the signal-to-noise ratio ranges from $\sim40$ in the lensed case, up to $\sim 120$ when delensing techniques are applied to the CMB anisotropy data. In these conditions, the identification of the $\nu\Lambda$CDM model would be definitely feasible. Moreover, the ability to distinguish the model from the massless case becomes achievable, even for the $M_{\nu}=0.06$ eV case.

\subsection{Realistic forecasts for future CMB Experiments and LSS Surveys}
\label{sec:7_real_results}
Although the ideal and the optimistic cases are certainly encouraging, the reality remains more challenging. This is because current state-of-the-art CMB and LSS experiments do not provide full-sky coverage and negligible noise effects. Not to mention the presence of foregrounds. 

For our analysis, we consider several LSS and CMB surveys: the Euclid photometric survey~\citep{EC20,Euclid_24} and LSST ``Gold'' and ``Optimistic'' samples~\citep{LSST_2009,LSST_2018,Gorecki_2014} as GC probes; the Euclid photometric survey as CS probe; SO~\cite{SO_2019}, CMB-S4~\citep{Abazajian_2016}, CMB-HD~\citep{HD_2019, CMBHD_2022, Macinnis_2024} and PICO~\citep{Hanany_2019, Aurlien_2023} as CMB experiments for CMBL and CMB-temperature. 
The specifications for the four CMB experiments considered are listed in Table~\ref{tab:so}, Table~\ref{tab:s4}, Table~\ref{tab:hd} and Table~\ref{tab:pico}.

In our realistic analysis, we test $5$ $\nu\Lambda$CDM models obtained from the baseline Planck 2018 cosmology~\cite{PlanckVI_2020}, generalised assuming $5$ different $M_{\nu}$ values: $0.06, 0.12, 0.18, 0.24, 0.30$ eV. We consider three scenarios of increasing complexity, taking into account the impact of CMB primary anisotropies, detector noise (D.N.), and residual foregrounds:
\begin{itemize}
    \item \label{itm:first} \textbf{Delen. + D.N.} : delensed $C_{\ell}^{TT}$ + detector noise;
    \item \label{itm:second}\textbf{Len. + D.N.} : lensed $C_{\ell}^{TT}$ + detector noise;
    \item \label{itm:third} \textbf{Len. + ILC} : lensed $C_{\ell}^{TT}$ + detector noise + residual foreground.
\end{itemize}

\begin{table}[!ht]
    \centering
    \begin{tabular}{|c|c|c|c|c|c|c|c|}
        \hline
        Frequency (GHz) \Tstrut 
        & \multicolumn{1}{c|}{27} 
        & \multicolumn{1}{c|}{39} 
        & \multicolumn{1}{c|}{93} 
        & \multicolumn{1}{c|}{145} 
        & \multicolumn{1}{c|}{225} 
        & \multicolumn{1}{c|}{280} \\[3pt] 
        \hline
        FWHM (arcmin) \Tstrut
        & \multicolumn{1}{c|}{7.4} 
        & \multicolumn{1}{c|}{5.1} 
        & \multicolumn{1}{c|}{2.2} 
        & \multicolumn{1}{c|}{1.4} 
        & \multicolumn{1}{c|}{1.0} 
        & \multicolumn{1}{c|}{0.9} \\[3pt] 
        \hline
        TT noise ($\mu$K-arcmin) \Tstrut
        & \multicolumn{1}{c|}{71} 
        & \multicolumn{1}{c|}{36} 
        & \multicolumn{1}{c|}{8.0} 
        & \multicolumn{1}{c|}{10} 
        & \multicolumn{1}{c|}{22} 
        & \multicolumn{1}{c|}{54} \\
        \hline
    \end{tabular}
    \caption{Experimental assumptions for SO. }
    \label{tab:so}
\end{table}

\begin{table}[!ht]
    \centering
    \begin{tabular}{|c|c|c|c|c|c|c|c|}
        \hline
        Frequency (GHz) \Tstrut 
        & \multicolumn{1}{c|}{20} 
        & \multicolumn{1}{c|}{27} 
        & \multicolumn{1}{c|}{39} 
        & \multicolumn{1}{c|}{93} 
        & \multicolumn{1}{c|}{145} 
        & \multicolumn{1}{c|}{225} 
        & \multicolumn{1}{c|}{280} \\[3pt] 
        \hline
        FWHM (arcmin) \Tstrut
        & \multicolumn{1}{c|}{10.0} 
        & \multicolumn{1}{c|}{7.4} 
        & \multicolumn{1}{c|}{5.1} 
        & \multicolumn{1}{c|}{2.2} 
        & \multicolumn{1}{c|}{1.4} 
        & \multicolumn{1}{c|}{1.0} 
        & \multicolumn{1}{c|}{0.9} \\[3pt] 
        \hline
        TT noise ($\mu$K-arcmin) \Tstrut
        & \multicolumn{1}{c|}{45.9} 
        & \multicolumn{1}{c|}{15.5} 
        & \multicolumn{1}{c|}{8.7} 
        & \multicolumn{1}{c|}{1.5} 
        & \multicolumn{1}{c|}{1.5} 
        & \multicolumn{1}{c|}{4.8} 
        & \multicolumn{1}{c|}{11.5} \\
        \hline
    \end{tabular}
    \caption{Experimental assumptions for CMB-S4. }
    \label{tab:s4}
\end{table}

\begin{table}[!ht]
    \centering
    \begin{tabular}{|c|c|c|c|c|c|c|}
        \hline
        Frequency (GHz) \Tstrut 
        & \multicolumn{1}{c|}{$30$} 
        & \multicolumn{1}{c|}{$40$} 
        & \multicolumn{1}{c|}{$90$} 
        & \multicolumn{1}{c|}{$150$} 
        & \multicolumn{1}{c|}{$220$} 
        & \multicolumn{1}{c|}{$280$} \\[3pt] 
        \hline
        FWHM (arcmin) \Tstrut
        & \multicolumn{1}{c|}{$1.30$} 
        & \multicolumn{1}{c|}{$0.94$} 
        & \multicolumn{1}{c|}{$0.41$} 
        & \multicolumn{1}{c|}{$0.25$} 
        & \multicolumn{1}{c|}{$0.17$} 
        & \multicolumn{1}{c|}{$0.13$} \\[3pt] 
        \hline
        TT noise ($\mu$K-arcmin) \Tstrut
        & \multicolumn{1}{c|}{$6.50$} 
        & \multicolumn{1}{c|}{$3.40$} 
        & \multicolumn{1}{c|}{$0.73$} 
        & \multicolumn{1}{c|}{$0.79$} 
        & \multicolumn{1}{c|}{$2.00$} 
        & \multicolumn{1}{c|}{$4.60$} \\
        \hline
    \end{tabular}
    \caption{Experimental assumptions for CMB-HD.}
    \label{tab:hd}
\end{table}

\begin{table}[!ht]

        \centering
        \begin{tabular}{|c|c|c|c|c|c|c|c|c|c|c|c|}
            \hline
            Frequency (GHz) \Tstrut 
            & \multicolumn{1}{c|}{21} 
            & \multicolumn{1}{c|}{25} 
            & \multicolumn{1}{c|}{30} 
            & \multicolumn{1}{c|}{36} 
            & \multicolumn{1}{c|}{43} 
            & \multicolumn{1}{c|}{52} 
            & \multicolumn{1}{c|}{62} 
            & \multicolumn{1}{c|}{75} 
            & \multicolumn{1}{c|}{90} 
            & \multicolumn{1}{c|}{108} \\[3pt] 
            \hline
            FWHM (arcmin) \Tstrut
            & \multicolumn{1}{c|}{38.4} 
            & \multicolumn{1}{c|}{32.0} 
            & \multicolumn{1}{c|}{28.3} 
            & \multicolumn{1}{c|}{23.6} 
            & \multicolumn{1}{c|}{22.2} 
            & \multicolumn{1}{c|}{18.4} 
            & \multicolumn{1}{c|}{12.8} 
            & \multicolumn{1}{c|}{10.7} 
            & \multicolumn{1}{c|}{9.5} 
            & \multicolumn{1}{c|}{7.9} \\[3pt] 
            \hline
            TT noise ($\mu$K-arcmin) \Tstrut
            & \multicolumn{1}{c|}{23.9} 
            & \multicolumn{1}{c|}{18.4} 
            & \multicolumn{1}{c|}{12.4} 
            & \multicolumn{1}{c|}{7.9} 
            & \multicolumn{1}{c|}{7.9} 
            & \multicolumn{1}{c|}{5.7} 
            & \multicolumn{1}{c|}{5.4} 
            & \multicolumn{1}{c|}{4.2} 
            & \multicolumn{1}{c|}{2.8} 
            & \multicolumn{1}{c|}{2.3} \\[3pt] 
            \hline
        \end{tabular}
    
    \vspace{1em} 

        \centering
        \begin{tabular}{|c|c|c|c|c|c|c|c|c|c|c|c|}
            \hline
            Frequency (GHz) \Tstrut
            & \multicolumn{1}{c|}{129} 
            & \multicolumn{1}{c|}{155} 
            & \multicolumn{1}{c|}{186} 
            & \multicolumn{1}{c|}{223} 
            & \multicolumn{1}{c|}{268} 
            & \multicolumn{1}{c|}{321} 
            & \multicolumn{1}{c|}{385} 
            & \multicolumn{1}{c|}{426}
            & \multicolumn{1}{c|}{555}
            & \multicolumn{1}{c|}{666} 
            & \multicolumn{1}{c|}{788} \\[3pt] 
            \hline
            FWHM (arcmin) \Tstrut
            & \multicolumn{1}{c|}{7.4} 
            & \multicolumn{1}{c|}{6.2} 
            & \multicolumn{1}{c|}{4.3} 
            & \multicolumn{1}{c|}{3.6} 
            & \multicolumn{1}{c|}{3.2} 
            & \multicolumn{1}{c|}{2.6} 
            & \multicolumn{1}{c|}{2.5} 
            & \multicolumn{1}{c|}{2.1} 
            & \multicolumn{1}{c|}{1.5} 
            & \multicolumn{1}{c|}{1.3} 
            & \multicolumn{1}{c|}{1.1} \\[3pt] 
            \hline
            TT noise ($\mu$K-arcmin) \Tstrut
            & \multicolumn{1}{c|}{2.1} 
            & \multicolumn{1}{c|}{1.8} 
            & \multicolumn{1}{c|}{4.0} 
            & \multicolumn{1}{c|}{4.5} 
            & \multicolumn{1}{c|}{3.1} 
            & \multicolumn{1}{c|}{4.2} 
            & \multicolumn{1}{c|}{4.5} 
            & \multicolumn{1}{c|}{9.1} 
            & \multicolumn{1}{c|}{45.8} 
            & \multicolumn{1}{c|}{177} 
            & \multicolumn{1}{c|}{1050} \\[3pt] 
            \hline
        \end{tabular}
\caption{Experimental assumptions for PICO (baseline).}
\label{tab:pico}
\end{table}

For the first two scenarios, the noise sensitivities of the different frequency channels are combined using an inverse-variance weighting, after convolution with the corresponding Gaussian beam window functions~\citep{Finelli_2018}:
\begin{equation}
    d_{\ell} = \left[ \sum_{\nu} \frac{1}{d_{\nu \, \ell}} \right]^{-1}\,.
\end{equation}
with
\begin{equation}
    d_{\nu \,  \ell} =  \Delta_{T \, \nu} ^{2} \, \text{exp}\left[ \frac{\theta_{FWHM \, \nu}^{2} \, \ell^{2}}{8 \, ln2} \right] \,.
\label{eq:dl}
\end{equation}
where $\Delta_{T \, \nu}$ is the noise level of the experiment (third row of Tables~\ref{tab:so}, ~\ref{tab:s4},~\ref{tab:hd},~\ref{tab:pico}) and $\theta_{FWHM \, \nu}$ is the full-width at half-maximum of a Gaussian beam (second row of Tables~\ref{tab:so},~\ref{tab:s4},~\ref{tab:hd},~\ref{tab:pico}).

For the third scenario, we use the Internal Linear Combination method~\citep{Bennett_2003b}, and specifically the one implemented in the \texttt{BasicILC} code\footnote{\url{https://github.com/EmmanuelSchaan/BasicILC}}, to optimally combine the different frequency channels, minimizing the contribution of noise and foregrounds. We account for the expected small-scale foregrounds at each frequency channel, specifically Sunyaev-Zel'dovich effects, Cosmic Infrared Background, and radio point sources. These components are modelled using the latest data from ground-based experiments such as the Atacama Cosmology Telescope (ACT,~\cite{ACT_2023}) and the South Pole Telescope (SPT,~\cite{SPT}). In this respect, our analysis is to be considered conservative and bound to be improved thanks to future higher-resolution measurements, for which it will be possible to more efficiently resolve and mask sources. Throughout our cross-correlation studies, we make the starting point approximation that any residual foreground in CMB anisotropy maps does not correlate with the tracers of the LSS, even in the case of CMBL. Finally, for ground-based experiments like SO, CMB-S4 and CMB-HD, we do not attempt to model the impact of the Earth's atmosphere, which predominantly affects 
large scales.

\begin{figure*}[!ht]
\centering
\includegraphics[width=0.65\textwidth]{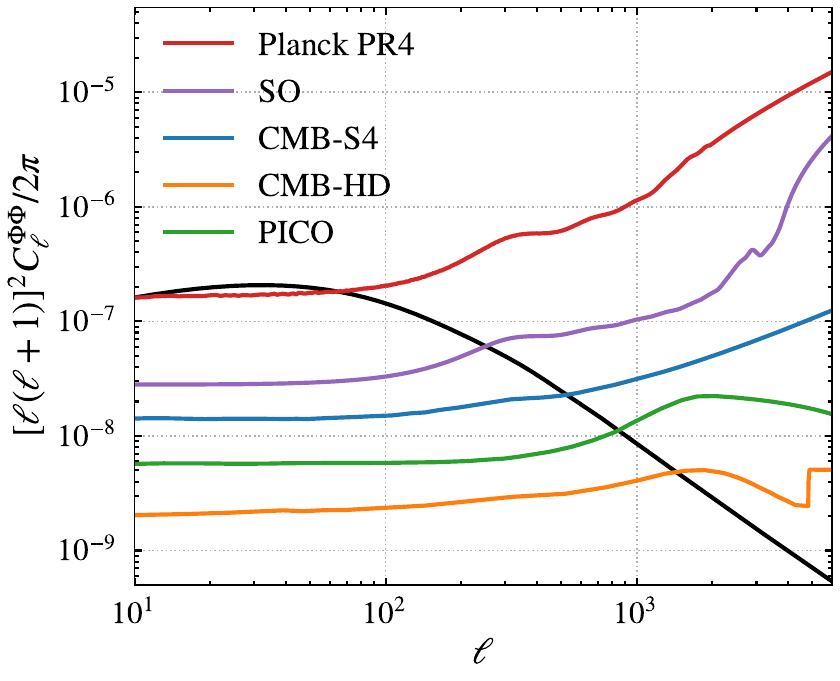} 
\caption{CMB-lensing potential reconstruction noise levels for the four CMB experiments tested: SO (purple), CMB-S4 (blue), CMB-HD (orange), PICO (green). The Planck PR4 (red) has been reported as a reference. In black the theoretical CMB lensing potential for the Planck cosmology.}
\label{fig:cmbl_noise}
\end{figure*}

In Equation~\eqref{eq:snr}, as noise for the gravitational potential tracers, $N_{\ell}$, we use the shot-noise $\Bar{n}^{-1}$ for the GC probes as given in Section~\ref{sec:cross-correlations}, the specific CMBL reconstruction noise for CMB-S4~\citep{EuclidXV_2022}, CMB-HD~\citep{Macinnis_2024} and PICO~\citep{Hanany_2019} (see Figure~\ref{fig:cmbl_noise}), while for the CS probe we use~\citep{Joachimi_2010}:
\begin{equation}
    N^{\gamma}_{\ell} = \frac{\sigma_{\epsilon}^{2}}{2 \, \Bar{n}} \,,
\end{equation}
where $\sigma_{\epsilon}^{2}$ is the variance of the intrinsic galaxy ellipticity, assumed to be $0.3$ for the Euclid photometric survey~\citep{EC20}.

Lastly, given the tested CMB and LSS experiments and their sky fraction ($f_{sky}$, Table~\ref{tab:fsky0}), we assume for our analysis the $f_{sky}$ listed in Table~\ref{tab:fsky}, as the result of their overlap~\citep{Allison_2015, Bermejo_2021, CMBHD_2022, CMB-S4, Ferraro_2022}.
\begin{table}[!ht]
\centering

\begin{minipage}{0.48\textwidth}
\centering
\begin{tabular}{|l|c|}
\hline
\multirow{1}{*}{}
& \multicolumn{1}{c|}{$f_{sky}$} \Tstrut\\[3pt] 
\hline
Euclid \Tstrut
& \multicolumn{1}{c|}{$0.36$} 
\\[3pt] 
\hline
LSST \Tstrut
& \multicolumn{1}{c|}{$0.40$} 
\\[3pt] 
\hline
SO \Tstrut
& \multicolumn{1}{c|}{$0.40$} 
\\[3pt] 
\hline
CMB-S4 \Tstrut
& \multicolumn{1}{c|}{$0.40$} 
\\[3pt] 
\hline
CMB-HD  \Tstrut
& \multicolumn{1}{c|}{$0.50$} 
\\[3pt] 
\hline
PICO  \Tstrut
& \multicolumn{1}{c|}{$0.90$} 
\\
\hline
\end{tabular}
\caption{Sky fraction of the tested LSS and CMB expertiments.}
\label{tab:fsky0}
\end{minipage}
\hfill
\begin{minipage}{0.48\textwidth}
\centering
\vspace{1.32cm}
\begin{tabular}{|l|c|c|}
\hline
\multirow{1}{*}{}
& \multicolumn{1}{c|}{Euclid} \Tstrut
& \multicolumn{1}{c|}{LSST}\Tstrut\\[3pt] 
\hline
SO \Tstrut
& \multicolumn{1}{c|}{$0.25$} 
& \multicolumn{1}{c|}{$0.40$}
\\[3pt] 
\hline
CMB-S4 \Tstrut
& \multicolumn{1}{c|}{$0.25$} 
& \multicolumn{1}{c|}{$0.40$}
\\[3pt] 
\hline
CMB-HD  \Tstrut
& \multicolumn{1}{c|}{$0.25$} 
& \multicolumn{1}{c|}{$0.40$}
\\[3pt] 
\hline
PICO  \Tstrut
& \multicolumn{1}{c|}{$0.36$} 
& \multicolumn{1}{c|}{$0.40$} 
\\
\hline
\end{tabular}

\caption{Overlapping sky fraction between galaxy surveys and CMB experiments.}
\label{tab:fsky}

\end{minipage}

\end{table}

To represent the results of the application of Equation~\eqref{eq:snr} and Equation~\eqref{eq:mnu} with all the aforementioned components, we use grouped bar charts. In the following Sections, Figures~\ref{fig:euclid_gc} - \ref{fig:cmbl}, for each cross-correlation investigated, the colour of the bars indicates the CMB experiment simulated to represent the ISWRS effect (or both the ISWRS and the $\Phi$-tracer in the case of the cross-correlation with CMBL). The shade of the colour indicates the $M_{\nu}$ model: we report the results for all the $\nu\Lambda$CDM cosmologies analysed, from lighter shades that represent lighter $M_{\nu}$, to darker that represent heavier models. 
The cumulative $S/N$ (left panels) and the $\sqrt{\chi^{2}}$ (right panels) represented by the bars correspond to the values reached in correspondence of $\ell_{\text{max}} = 6000$. 

\subsection{Galaxy clustering -- Optimal Weighting}
The cross-correlation between the ISWRS effect and GC is the most widely investigated, primarily because of the higher cross-angular power spectrum signal. Several techniques have also been developed to reshape the galaxy number count distribution in order to optimize the S/N~\citep{Alonso_2020, Urban_2021,Ferraro_2022}. For this cross-correlation only, in fact, together with the three scenarios listed in the previous Section, we consider even a fourth scenario. Following the example of Ferraro et al. (2022)~\cite{Ferraro_2022}, we implement a method to improve the overlap between ISWRS and GC window functions and to maximise the $S/N$: the optimal weighting (O.W.).

\paragraph{O.W.}
We know that the majority of galaxy surveys are characterized by a large number of galaxies at low redshift ($z \sim 1$), but the peak of the ISWRS is at high redshift ($ z \sim 2-4$). To maximise the overlap between the window functions, and consequently maximise the $S/N$, we can up-weight galaxies at higher redshift via the usage of~\cite{Ferraro_2022}: 
\begin{equation}
    w(z) \propto \frac{b(z)\, W^{\dot\Phi}(k^{*},z)\,P_{\delta\delta}(k^{*},z)}{\left(\frac{n(z)}{\Bar{n}}\right) \, \left[ b^{2}(z) \, P_{\delta\delta}(k^{*},z) + \frac{\chi^{2}}{H(z)\,n(z)} \right]} \,,
\label{eq:ow}
\end{equation}
which is the expression of the O.W., where $k^{*} = \frac{\ell^{*} + 1/2}{\chi(z)}$, with $\ell^{*} = 5000$, as this is the scale from which most of the ISWRS signal comes from~\cite{Ferraro_2022}. 

\begin{figure*}[!ht]
\centering
\includegraphics[width=0.99\textwidth]{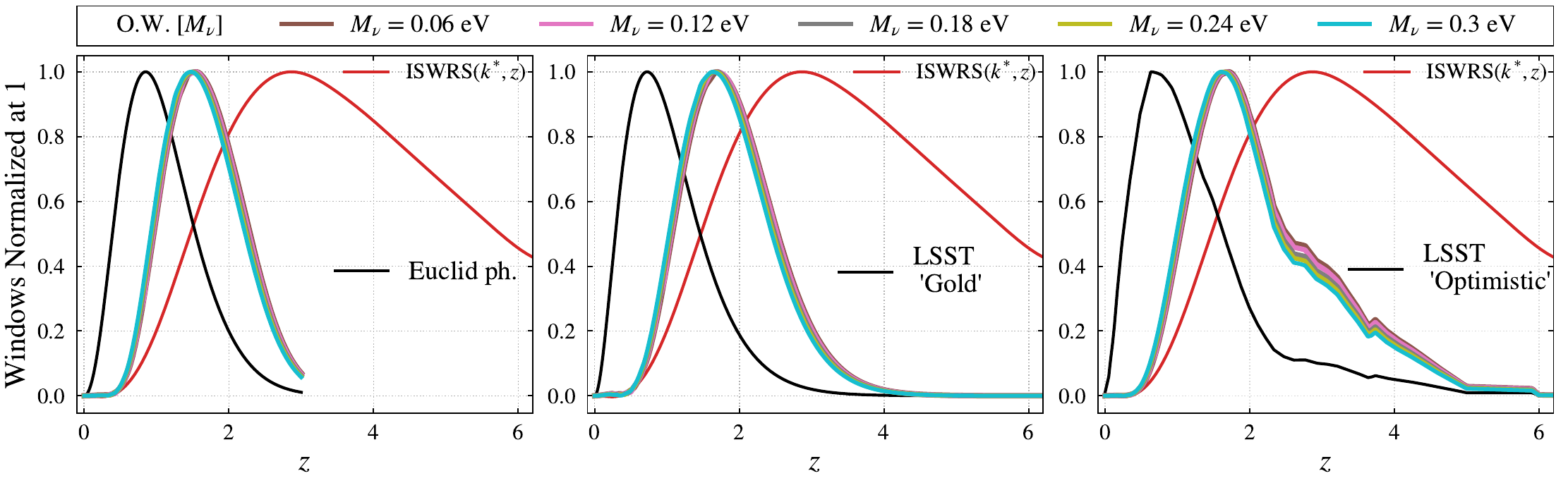}
\caption{Window functions of the photometric Euclid, LSST ``Gold'' and LSST ``Optimistic'' GC samples from left to right (black solid lines), together with the ISWRS response as a function of $z$ computed at a scale $k^{*} = (\ell^{*} + 1/2)/\chi(z)$, where $\ell^{*} = 5000$ (red lines). In each panel, the results for the application of the optimal weighting (O.W.) to each survey have been reported. Being the O.W. cosmology-dependent, we report the results obtained for each of the $\nu\Lambda$CDM cosmologies tested, with $M_{\nu}$ = $0.0$ eV (cyan), $0.06$ eV (brown), $0.12$ eV (pink), $0.18$ eV (grey), $0.24$ eV (acid green), $0.30$ eV (cyan).}
\label{fig:ow_windows}
\end{figure*}

Transforming then $W^{g}(k,z)$ into $w(k^{*},z) \, \cdot \, W^{g}(k,z)$ increases the overlap between the window functions, as illustrated in Figure~\ref{fig:ow_windows}. This comes at the cost of increasing the shot-noise, because higher redshift galaxies are rare. It is converted from the usual $\Bar{n}^{-1}$ to~\cite{Ferraro_2022, Alonso_2020, Urban_2021}:
\begin{equation}
    \frac{1}{\Bar{n}_{O.W.}} = \frac{1}{\Bar{n}^{2}} \, \int {\rm d}z \, n(z) \, w^{2}(z) \,.
\end{equation}
However, since Equation~\eqref{eq:ow} is the result of the minimisation of the noise via the Lagrangian multiplier, this construction allows to balance the shot-noise increase, as comprehensively explained in Appendix A of~\cite{Ferraro_2022}.

As Equation~\eqref{eq:ow} depends on $P_{\delta\delta}(k,z)$, in our forecast we consistently derive a specific $w(k^{*},z)$ for each $\nu\Lambda$CDM cosmology tested. This is the reason why in Figure~\ref{fig:ow_windows} there is a different $w(k^{*},z) \, W^{g}(k,z)$ for each $M_{\nu}$ value, showed for $k = k^{*}$. 

In a real data analysis, before forming the galaxy overdensity map, each galaxy in the catalogue should be weighted by $w(z)$ in Equation~\eqref{eq:ow}, based on its photometric redshift. However, the underlying cosmological model is not known \textit{a priori}. A viable strategy would be to adopt an iterative approach: start from a fiducial cosmological model, infer the cosmological parameters from the data, recompute the weights using the best-fitting model, and repeat the analysis until convergence is reached. We also note that in this work we neglect photometric redshift uncertainties, since the weights are smooth functions of redshift and we do not attempt a tomographic analysis. This approximation is expected to be adequate unless the catalogue contains a significant fraction of catastrophic outliers. The optimisation and validation of the weighting scheme for real data are left to future work.

\subsubsection{Euclid photometric (GC) $\times$ ISWRS}
\begin{figure*}[!ht]
\centering
\hspace{-0.83cm}
\begin{tabular}{cc}
\includegraphics[width=0.49\textwidth]{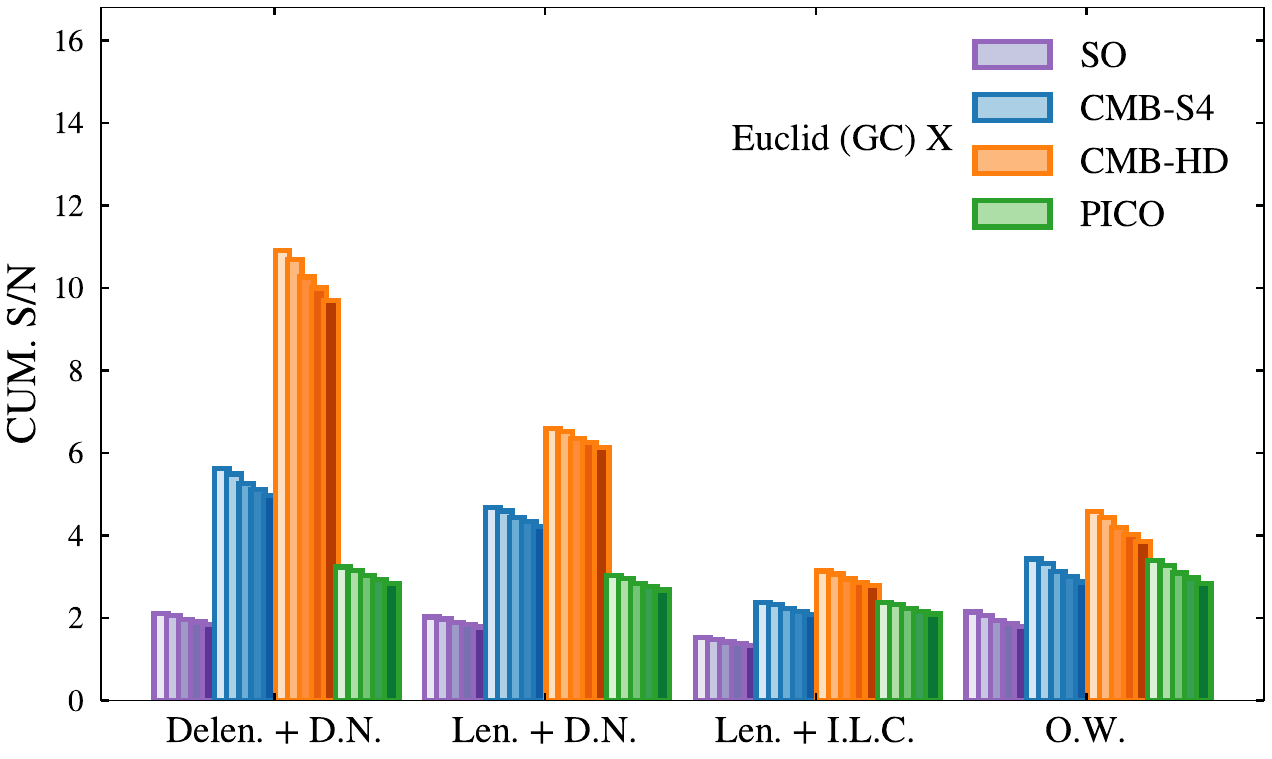} &
\includegraphics[width=0.503\textwidth]{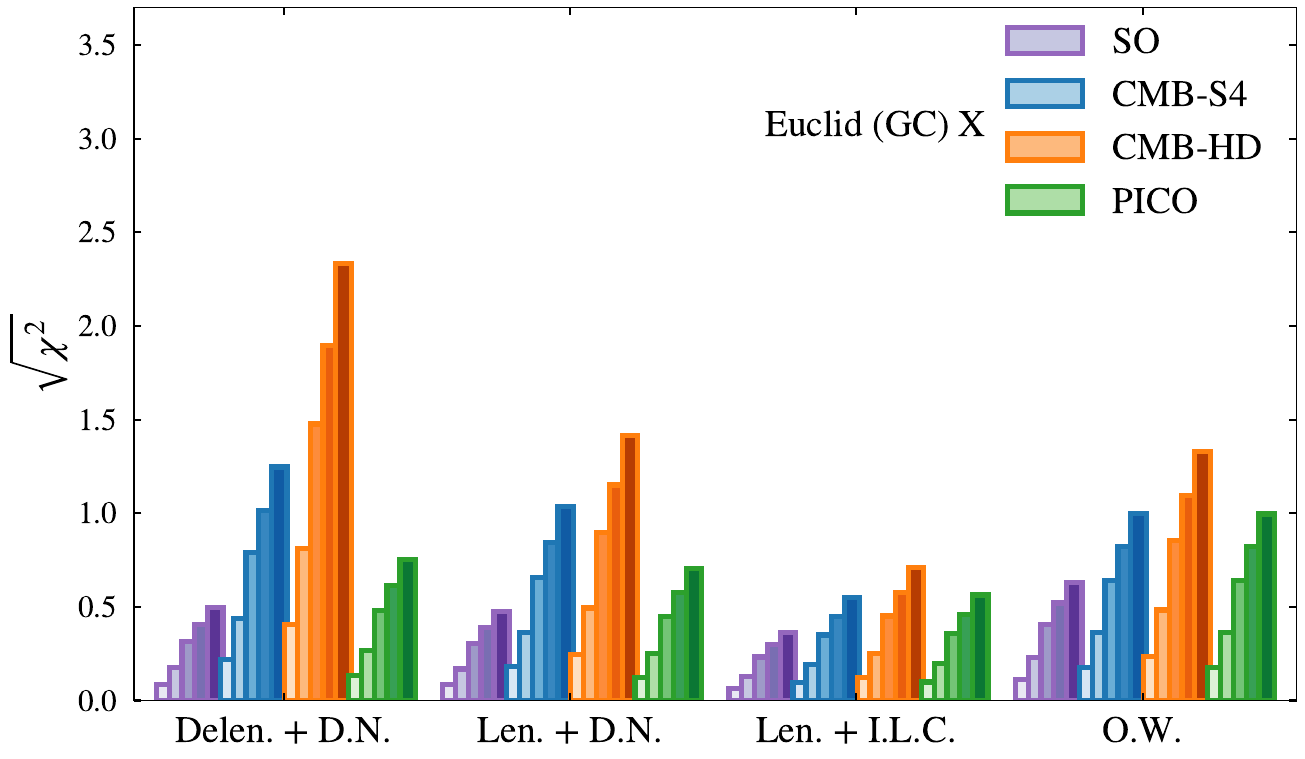}
\end{tabular}
\caption{Results for the photometric Euclid GC survey cross-correlation with all the CMB experiments, for all the $\nu\Lambda$CDM models analysed. For each of the four scenarios tested, the purple bars represent the results for the cross-correlation with SO, the blue bars for those with CMB-S4, the orange bars for those with CMB-HD and the green bars for those with PICO. For every combination of the Euclid (GC) survey with a CMB experiment, the results for all the $5$ $\nu\Lambda$CDM models have been represented with a bar (with $M_{\nu}$ increasing from $0.06$ eV to $0.30$ eV from left to right).
The left panel shows the results of the cumulative $S/N$ in correspondence of $\ell_{max}=6000$ computed with Equation~\eqref{eq:snr}. The right one shows the possibility of distinguishing the $\nu\Lambda$CDM model with respect to the massless case, computed with Equation~\eqref{eq:mnu}.}
\label{fig:euclid_gc}
\end{figure*}

Figure~\ref{fig:euclid_gc} shows the results for the combination of the photometric Euclid GC survey with the four CMB experiments. In the left panel, we represent the cumulative S/N results from the application of Equation~\eqref{eq:snr}. 
A detection of the ISWRS effect appears feasible with next-generation CMB experiments, provided that foregrounds can be kept under control or that optimal weighting of the GC is efficiently applied.
The highest detections are achieved when considering delensed $C_{\ell}^{TT}$ affected only by detector noise, particularly for the combination with CMB-HD, where we find an $\sim 11\sigma$ detection for $M_{\nu} = 0.06$ eV. The lowest S/N level occurs when considering lensed $C_{\ell}^{TT}$ combined with detector noise and foregrounds, especially in the case of the combination with SO where there is a $\sim 1.2\sigma$ detection for $M_{\nu}= 0.30$ eV. We want then to highlight the recovery achieved when applying the O.W. to the case of lensed $C_{\ell}^{TT}$ + ILC. The improvement in the detection is of $\sim38 \%$ in the combination with SO, $\sim41 \%$  with CMB-S4, $\sim42 \%$  with CMB-HD and $\sim39 \%$ with PICO.
In the right panel, we present the results from the application of Equation~\eqref{eq:mnu}, which provides information on the ability to distinguish a specific $\nu\Lambda$CDM model from the massless case. 
Only for CMB-HD, in the most optimistic conditions, there is the possibility to set constraints on the neutrino mass. Considering the current constraints on $M_{\nu}$ ($<0.11$ eV $95\%$ C.L.;~\cite{Allali_2024}), this cross-correlation analysis, with these combinations of experiments, will not lead to new constraints on neutrino masses.  

\subsubsection{LSST ``Gold'' (GC) $\times$ ISWRS}
\begin{figure*}[!ht]
\centering
\hspace{-0.83cm}
\begin{tabular}{cc}
\includegraphics[width=0.49\textwidth]{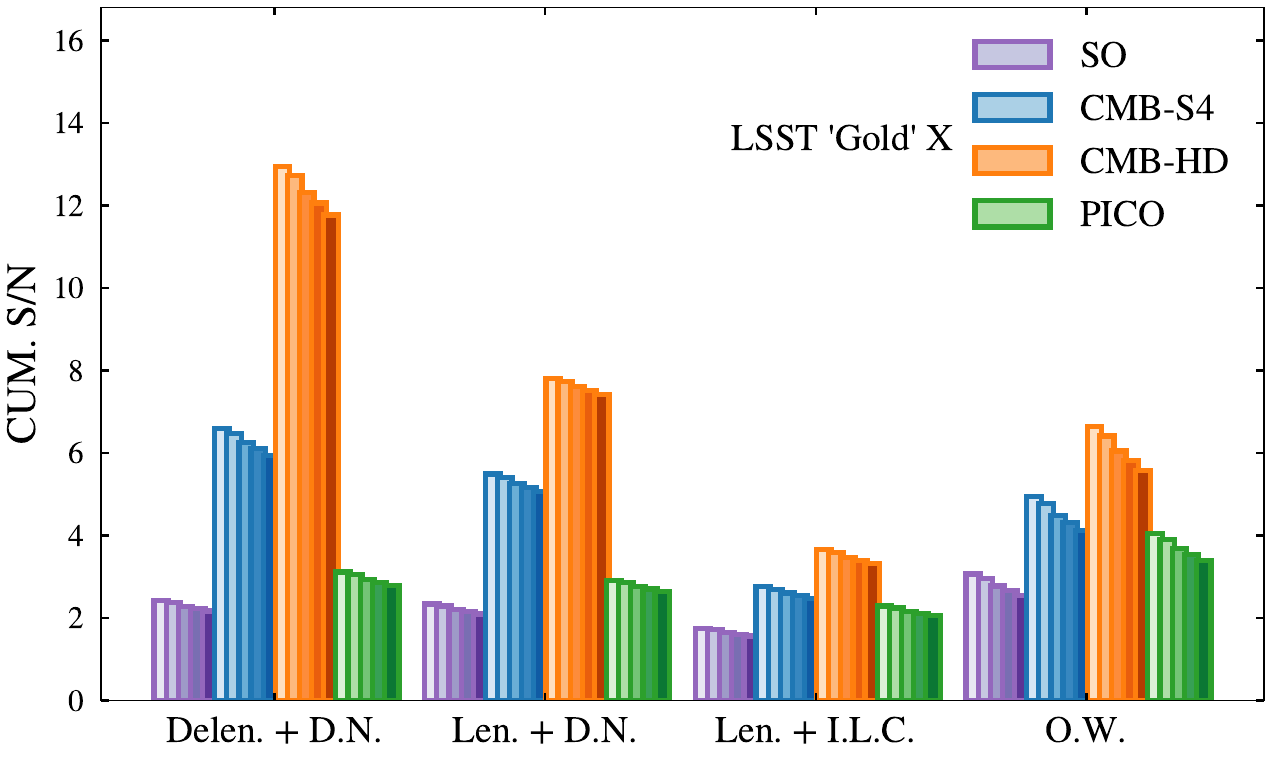} &
    \includegraphics[width=0.503\textwidth]{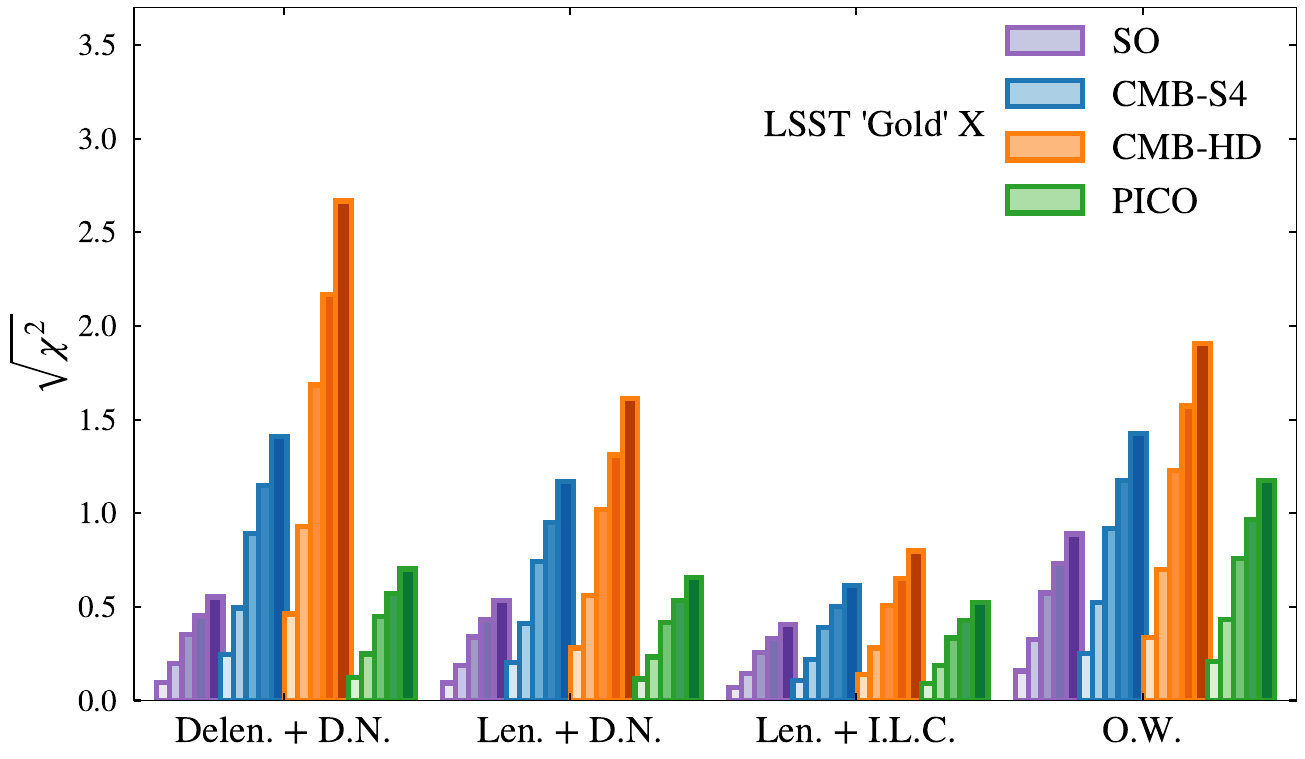}
\end{tabular}
\caption{Same as Figure~\ref{fig:euclid_gc}, but for the LSST ``Gold'' sample.}
\label{fig:lsst_G_gc}
\end{figure*}

Figure~\ref{fig:lsst_G_gc} shows the results for the combination of the LSST ``Gold'' sample with the four CMB experiments. In this case, the trend of the bars of the left panel is very similar to the one of Figure~\ref{fig:euclid_gc}. However, because of the higher $\Bar{n}$ and  the larger sky coverage overlap with the CMB experiments, the S/N levels achieved with the LSST ``Gold'' sample are overall higher. 
Our results reach a maximum of $\sim 12\sigma$ for the combination with CMB-HD, when considering delensed $C_{\ell}^{TT}$ affected only by detector noise for $M_{\nu} = 0.06$ eV, and a minimum of $\sim 1.5\sigma$ in the combination with CMB-SO, when considering lensed $C_{\ell}^{TT}$ combined with detector noise and foreground for $M_{\nu} = 0.30$ eV. Even in this case, the large recovery when applying the O.W. is evident, with an increase of the S/N level of $\sim69 \%$ for the combination with SO, $\sim73 \%$ with CMB-S4, $\sim75 \%$  with CMB-HD and $\sim71 \%$ with PICO. The results in the right panel of Figure~\ref{fig:lsst_G_gc} confirm the improvement obtained with these combinations of experiments.
Again the most promising results are obtained with CMB-HD, with some discriminating power at the $1.5\sigma$ level also for CMB-S4. Nevertheless, these promising results are achieved for $M_{\nu}$ values higher than $0.11$ eV ($95\%$ C.L.~\cite{Allali_2024}) and make also these combinations not constraining.

\subsubsection{LSST ``Optimistic'' (GC) $\times$ ISWRS}
\begin{figure*}[!ht]
\centering
\hspace{-0.83cm}
\begin{tabular}{cc}
\includegraphics[width=0.49\textwidth]{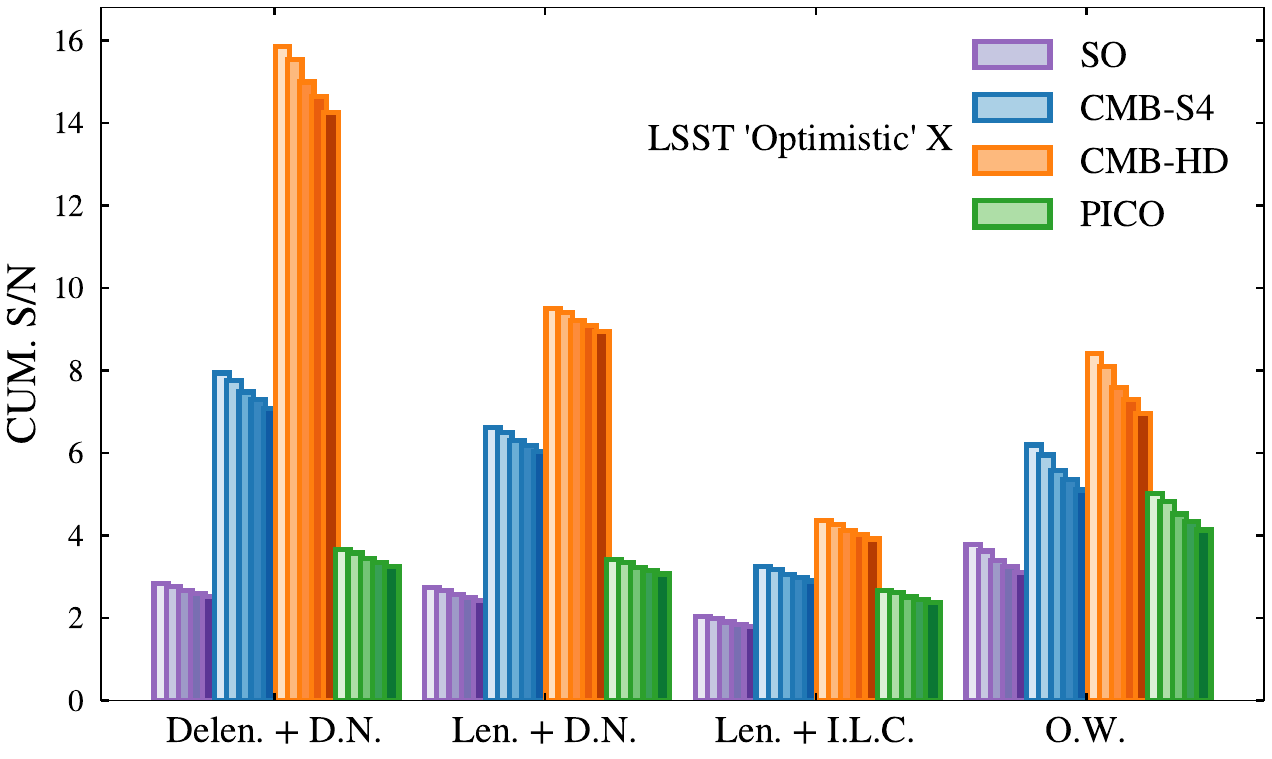} &
    \includegraphics[width=0.503\textwidth]{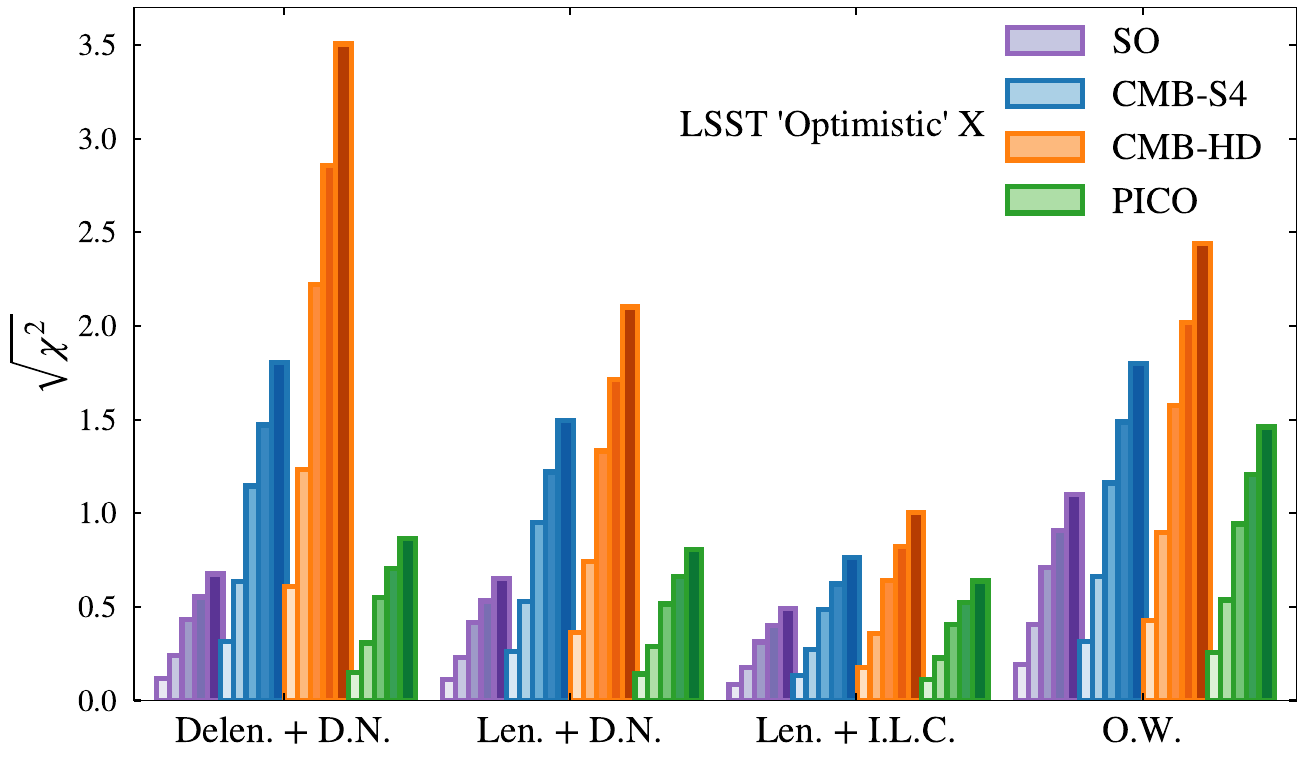}
\end{tabular}
\caption{Same as Figure~\ref{fig:euclid_gc}, but for the LSST ``Optimistic'' sample.}
\label{fig:lsst_F_gc}
\end{figure*}

Figure~\ref{fig:lsst_F_gc} shows the best results achieved for the cross-correlation of GC with ISWRS, obtained from the combination of the LSST ``Optimistic'' sample with the four CMB experiments. This was expected from the superposition of the window functions (both with and without the O.W. application) shown in the right panel of Figure~\ref{fig:ow_windows}, and from the very high $\Bar{n}$ (i.e.,  very low shot-noise). Here not only a minimum of $2\sigma$ detection level is always guaranteed, but a $7\sigma$ detection level is reached for all the cosmologies tested, in the first, the second and the fourth scenarios, in combination with CMB-HD. Achieving this result for all the analysed $\nu\Lambda$CDM cosmologies when applying the O.W. suggests the huge potential that this technique can have on real measurements in general, and on LSST ones in particular.
Overall, the results achieved range from a maximum of $\sim 16\sigma$ detection for the combination with CMB-HD, when considering delensed $C_{\ell}^{TT}$ affected only by detector noise for $M_{\nu} = 0.06$ eV, and a minimum of $\sim 2\sigma$ detection in the combination with SO, when considering lensed $C_{\ell}^{TT}$ combined with detector noise and residual foregrounds for $M_{\nu} = 0.30$ eV. The recovery here when applying the O.W. is of $\sim78 \%$ in combination with SO, $\sim83 \%$ with CMB-S4, $\sim85 \%$  with CMB-HD and $\sim81 \%$ with PICO, certainly larger with respect to the previous cases. The results for the $M_{\nu}$ investigation are more interesting, as we reach $>1\sigma$ detection level for $M_{\nu}=0.12$ eV in combination with CMB-HD in the first scenario. Although the conditions of the first scenario are undoubtedly idealistic and not achievable in the near future, they make this cross-correlation, with this combination of experiments, a promising probe to infer tighter constraints on $M_{\nu}$. 

\subsection{Cosmic Shear}
\label{sec:cross-cs}
The superposition between the window functions of CS and the ISWRS effect is significantly lower with respect to the case of GC (see Figure~\ref{fig:windows}). This results in a lower signal for the cross-correlation of these two fields, as shown in the middle panel of Figure~\ref{fig:cross-spectra}, that, together with a noise of the order of the GC one, makes this detection more challenging. Moreover, the analytical reconstruction of this cross-correlation is less precise when compared to the others, due to the greater sensitivity of the CS field to nonlinearities. Because of the assumptions made in the \texttt{Halofit} model, the analytical implementation does not provide a correct recovery of the amplitude of this cross-correlation in the nonlinear regime, and in particular  underestimates the results of N-body simulations (see Appendix~\ref{app:validation}), negatively impacting our analysis.

\subsubsection{Euclid photometric (CS) $\times$ ISWRS}
\begin{figure*}[!ht]
\centering
\hspace{-0.83cm}
\begin{tabular}{cc}
\includegraphics[width=0.49\textwidth]{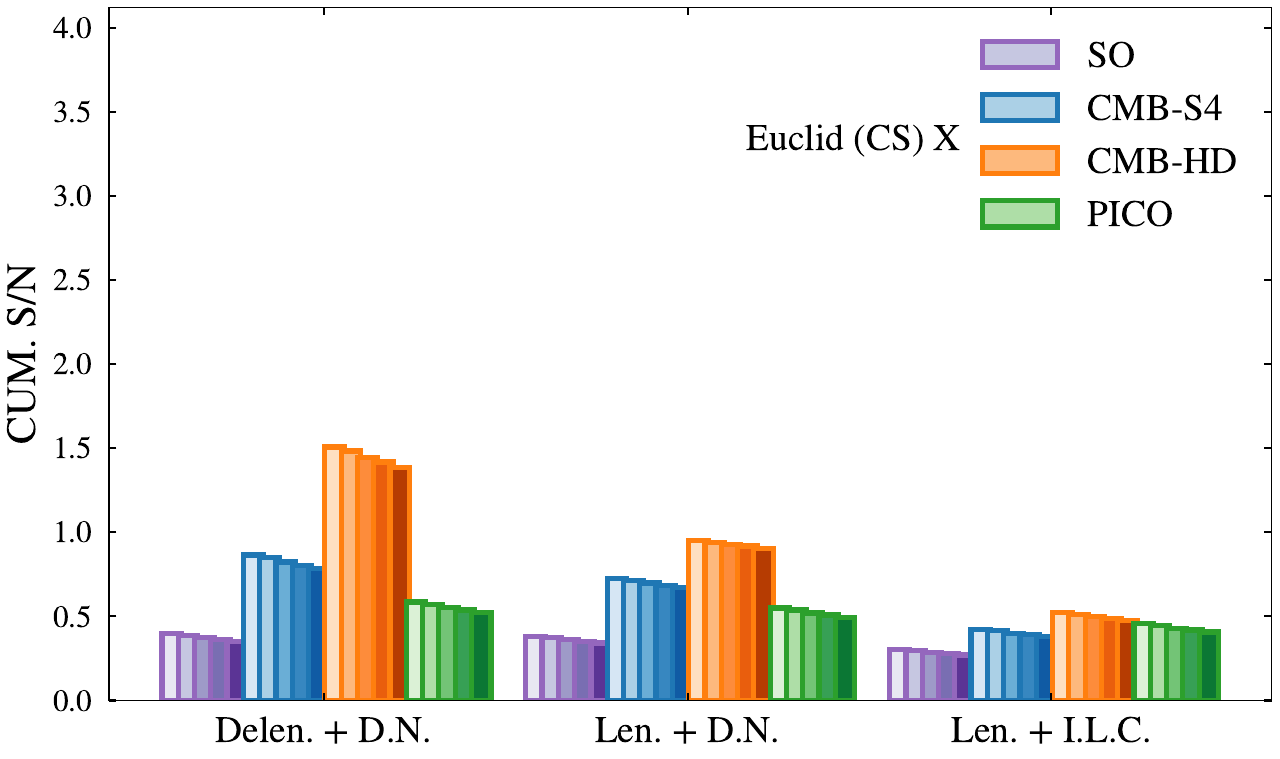} &
    \includegraphics[width=0.503\textwidth]{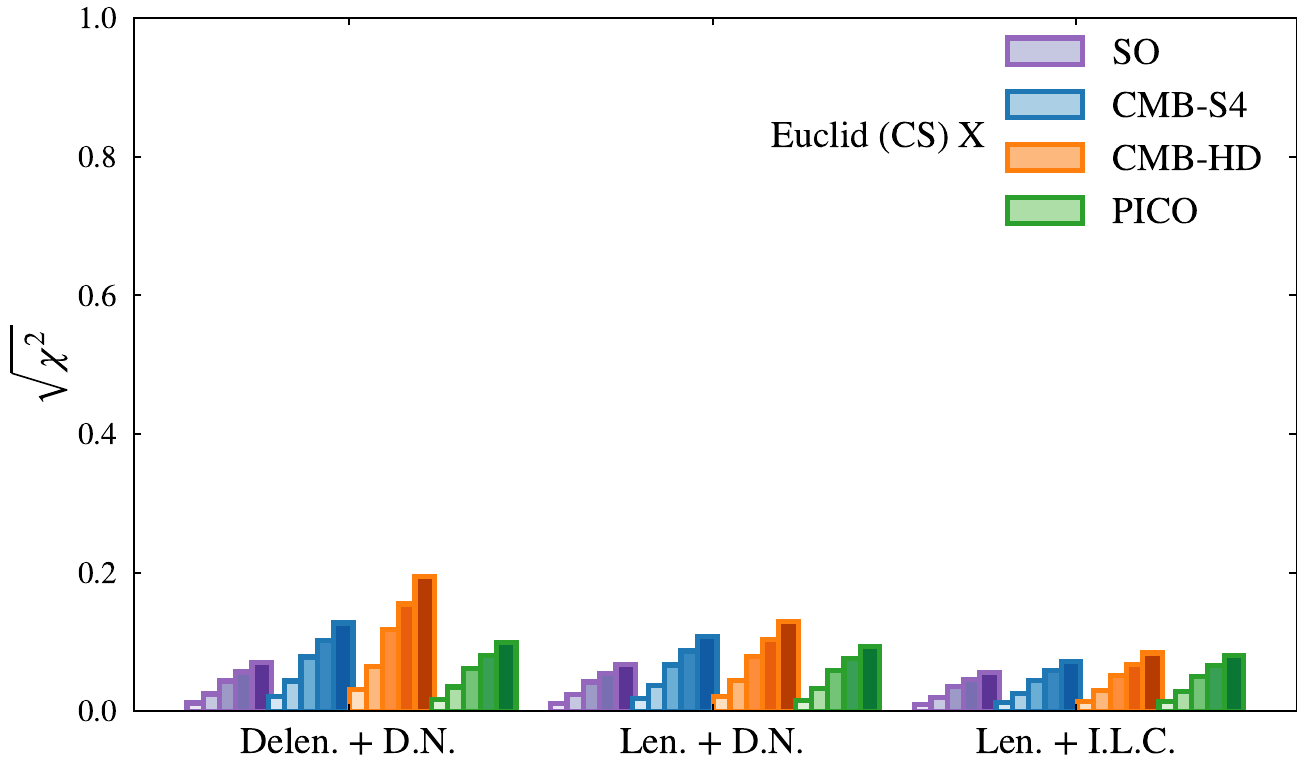}
\end{tabular}
\caption{Same as Figure~\ref{fig:euclid_gc}, but for the Euclid photometric CS combined with the four CMB experiments, for the three scenarios tested in this case.}
\label{fig:euclid_cs}
\end{figure*}

Figure~\ref{fig:euclid_cs} shows the results for the combination of the photometric Euclid CS survey with the four CMB experiments. It is immediately evident that these results are the less promising findings of this work. A $\gsim 1\sigma$ level is reached only in the case of the combination with CMB-HD, in the first and most optimistic scenario, with all the $\nu\Lambda$CDM cosmologies tested. The highest S/N level is achieved within this case and corresponds to $\sim 1.5\sigma$ for $M_{\nu}= 0.06$ eV.
Not guaranteeing the detection, this cross-correlation does not provide any chance at all to identify the cosmological model, as shown in the right panel, where the results barely reach the $0.2\sigma$ level.

\subsubsection{N-body simulation analysis}
\label{sec:nbody}
Since our analytical implementation does not provide a correct recovery of the amplitude of this cross-correlation, we perform a comparison between the detectability results obtained with our analytic implementation and those obtained using the spectra extracted from N-body simulations. Specifically, for the latter case, we use the cross-spectra extracted from a set of ISWRS and CS correlated DEMNUni maps, realised for the Planck 2014b cosmology~\cite{PlanckXVI_2014}, extended to the cases where $M_{\nu} = 0.0, 0.17. 0.30, 0.53$ eV, in the redshift range $z = [0.2,2.5]$.  The CS maps have been produced using the techniques described in~\cite{Carbone_2008, Calabrese_2015, Fabbian_2018}, assuming the galaxy distribution given by Equation~\eqref{eq:euclid_n}. For a proper comparison, the same cosmologies and redshift range have been adopted to reassess our analytical results.

\begin{figure*}[!ht]
\centering
\hspace{-0.825cm}
\begin{tabular}{cc}
\includegraphics[width=0.49\textwidth]{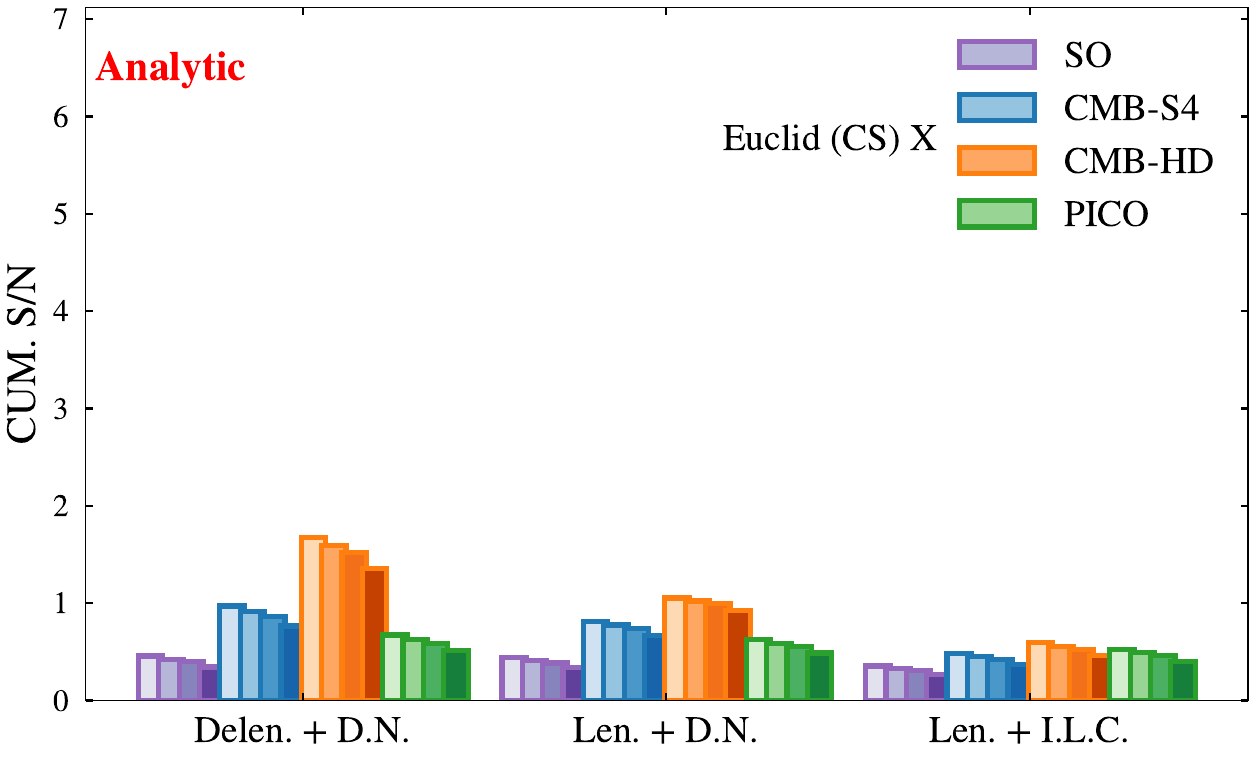} &
\includegraphics[width=0.503\textwidth]{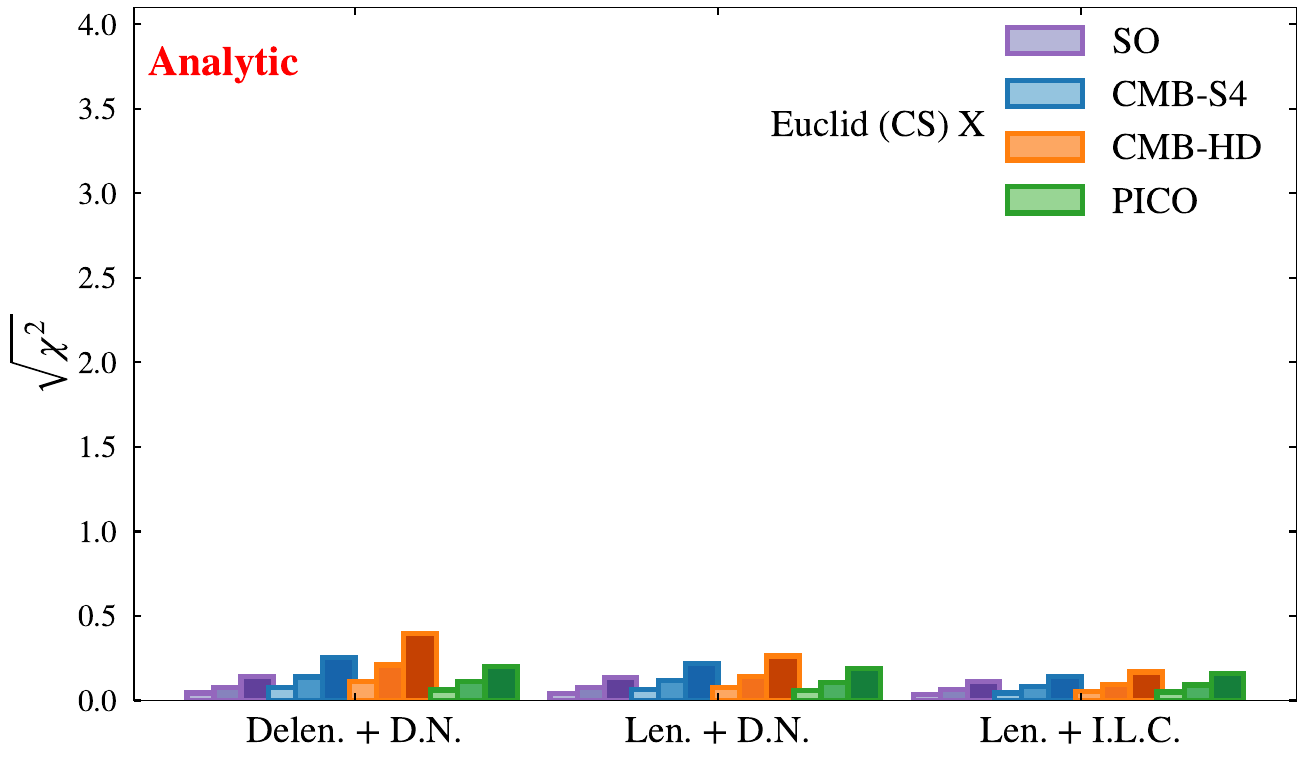} \\

\includegraphics[width=0.49\textwidth]{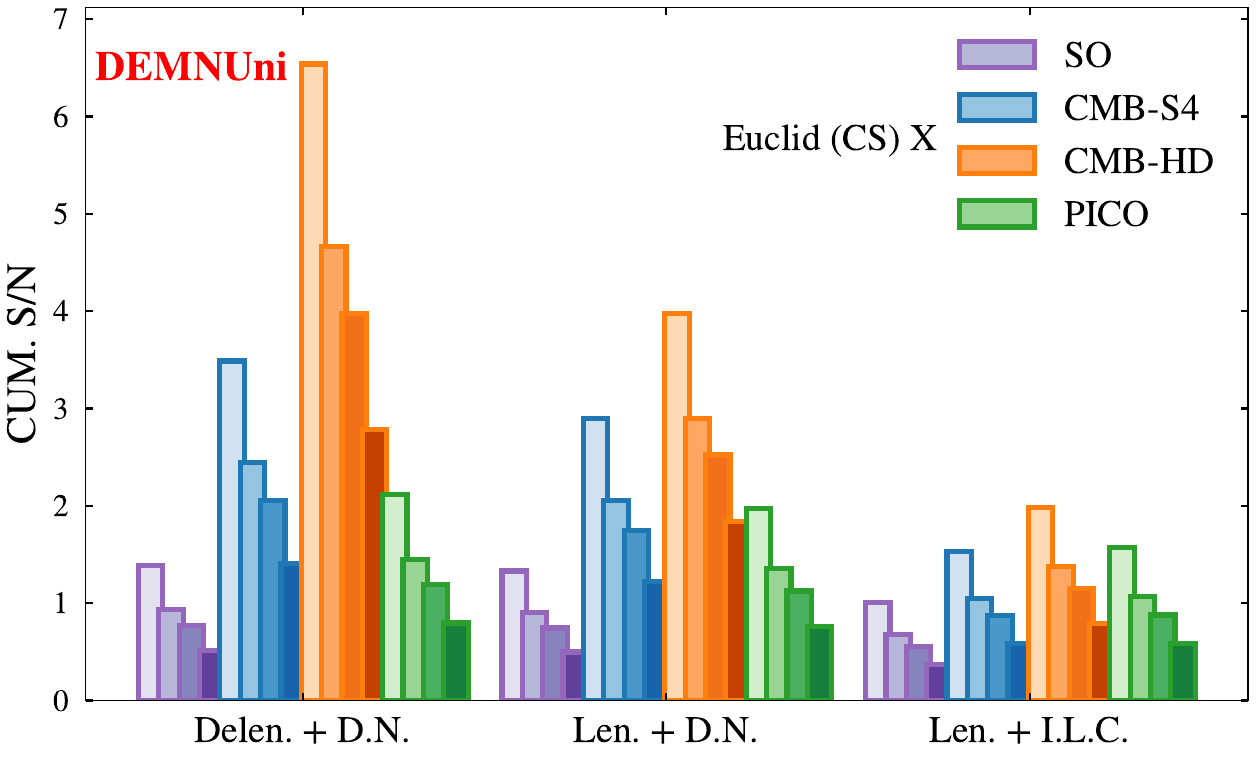} &
\includegraphics[width=0.503\textwidth]{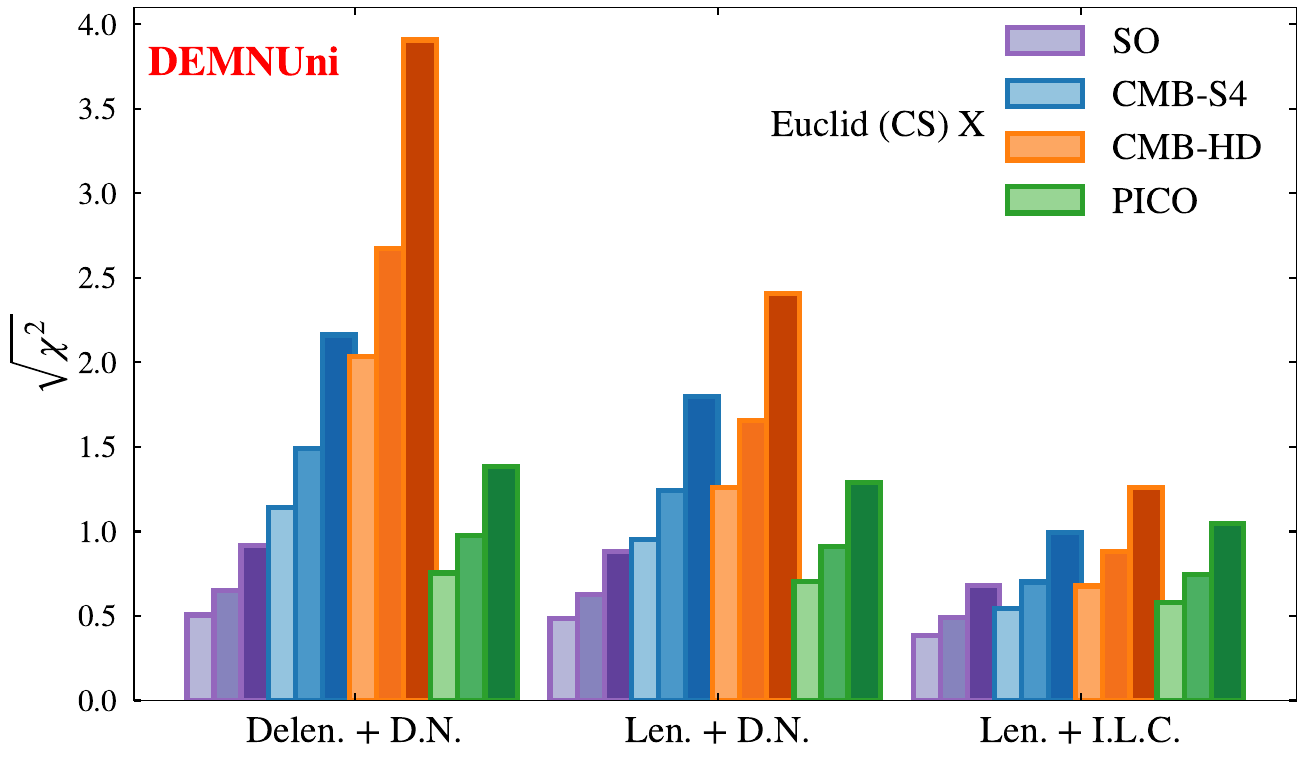}
\end{tabular}
\caption{\textit{Upper panels}: Same as Figure~\ref{fig:euclid_cs}, but realised for the case of a massless neutrino Planck 2014b cosmology~\cite{PlanckXVI_2014}, generalised to four $\nu\Lambda$CDM cases with $M_{\nu} = 0.0, 0.17, 0.30. 0.53$ eV.  
\textit{Lower panels}: Same as the upper panels, but obtained using the cross-spectra extracted from the DEMNUni maps realised for the same cosmologies.}
\label{fig:nbody}
\end{figure*}

It is clear from Figure~\ref{fig:nbody} that using the results of N-body simulations significantly improves the S/N ratio, leading to a maximum of $6.5\sigma$ for the $M_{\nu}=0.6$ eV $\nu\Lambda$CDM cosmology, in the case of the combination with CMB-HD for delensed $C_{\ell}^{TT}$ assuming only detector noise, and to a minimum level of $\sim 0.3\sigma$ in combination with SO, when considering lensed $C_{\ell}^{TT}$ with detector noise and foregrounds for $M_{\nu} = 0.30$ eV. These same configurations computed analytically result in a S/N level that ranges from a maximum of $\sim1.6\sigma$ in the combination with CMB-HD for delensed $C_{\ell}^{TT}$ assuming only detector noise, to a minimum of $\sim 0.2\sigma$ in the combination with SO, when considering lensed $C_{\ell}^{TT}$ with both detector noise and foregrounds for $M_{\nu} = 0.30$ eV. Accordingly to the improvement in the detection level, the possibility to disentangle the $M_{\nu}$ value is restored when using N-body simulations, whence a $\gsim 1\sigma$ level is reached with: CMB-S4 in the first (for $M_{\nu} = 0.17, 0.30, 0.53$ eV) and in the second (for $M_{\nu} = 0.30, 0.53$ eV) scenarios; CMB-HD in the first (for $M_{\nu} = 0.17, 0.30, 0.53$ eV), the second (for $M_{\nu} = 0.17, 0.30, 0.53$ eV) and in the third (for $M_{\nu} = 0.53$ eV) scenarios; PICO in the first (for $M_{\nu} = 0.53$ eV), the second (for $M_{\nu} = 0.53$ eV) and in the third (for $M_{\nu} = 0.53$ eV) scenarios.

These results indicate that with advancements in nonlinear modelling, substantial enhancements in the characterisation of this cross-correlation can be obtained. Such improvements would enable results comparable to those achieved with GC and CMBL for the detection of the ISWRS and the identification of $M_{\nu}$.

\subsection{CMB-Lensing}
The investigation of the cross-correlation between CMBL and the ISWRS effect has on the one hand the advantages of a larger sky fraction and of an extremely larger superposition between the window functions (Figure \ref{fig:windows}), while on the other hand the CMBL comes with the cost of a non-negligible reconstruction noise, that significantly affects this detection. However, the next generation CMB experiments aim to achieve unprecedented noise levels, as illustrated in Figure~\ref{fig:cmbl_noise}, which are expected to greatly enhance detection capabilities.

\subsubsection{CMBL $\times$ ISWRS} 
\begin{figure*}[!ht]
\centering
\hspace{-0.83cm}
\begin{tabular}{cc}
\includegraphics[width=0.49\textwidth]{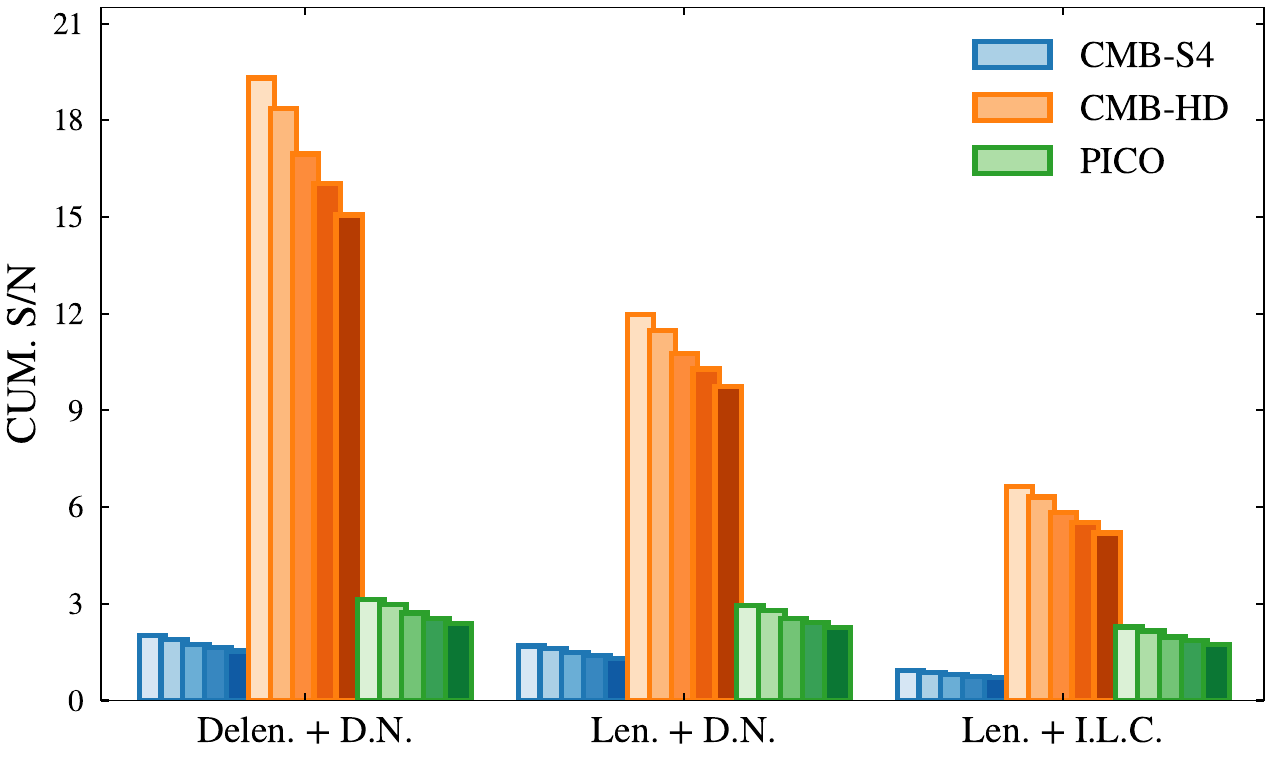} &
    \includegraphics[width=0.503\textwidth]{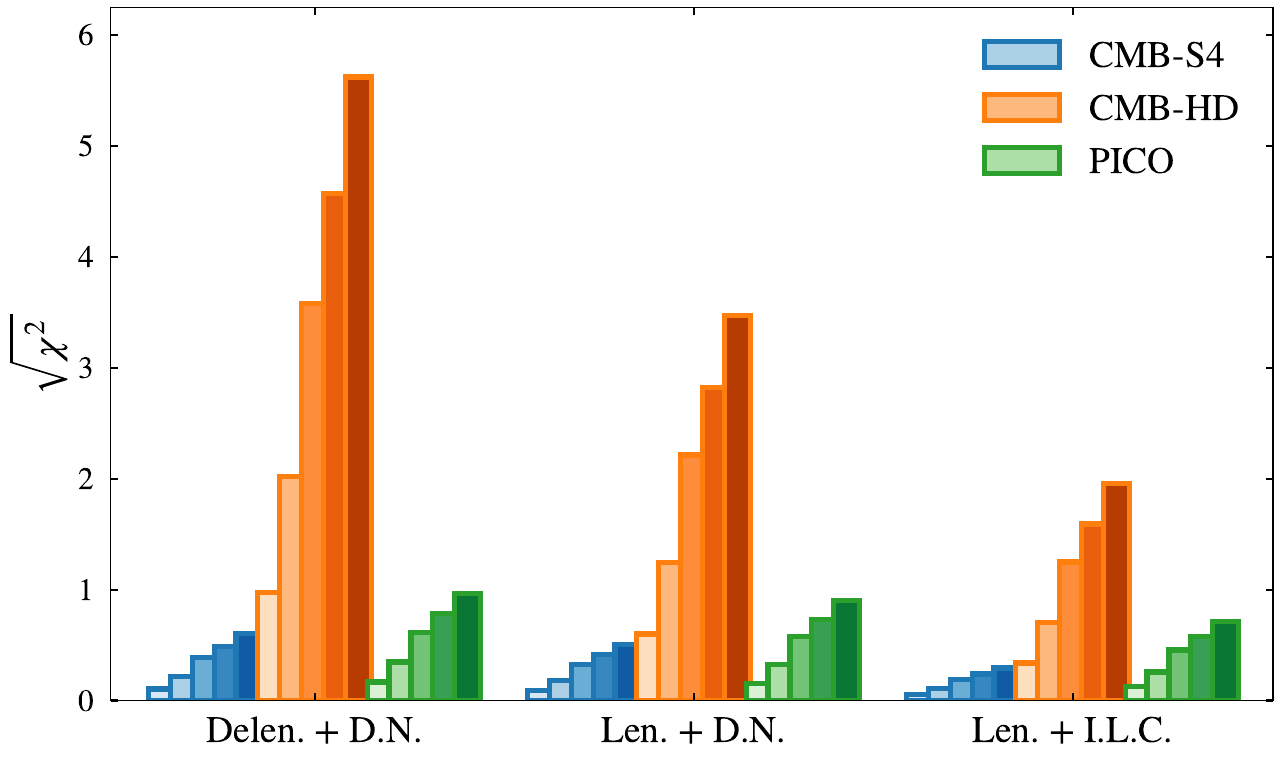}
\end{tabular}
\caption{Same as Figure~\ref{fig:euclid_gc}, but for the case of the cross-correlation between CMBL and the ISWRS effect, with both fields measured with CMB-S4, CMB-HD and PICO, for the three scenarios tested in this case. Results for SO are not shown because of the low forecasted S/N.}
\label{fig:cmbl}
\end{figure*}

Figure~\ref{fig:cmbl} shows the most outstanding results from this study. This is due to the fact that the CMBL reconstruction noise of CMB-HD is significantly lower than that of SO, CMB-S4 and PICO, with a marked reduction at the scales relevant for the ISWRS effect. As a result, its measurement of CMBL makes it the most effective $\Phi$-probe among those tested in this work. In this figure, we do not show results for SO, as the corresponding CMBL reconstruction noise level results in low S/N values for the cross-correlation. On the other hand, the results achieved when mocking CMB-HD significantly overcome a $5\sigma$ detection level in all the cosmologies tested in the cases of lensed and delensed $C_{\ell}^{TT}$, with and without I.L.C. This makes CMB-HD a perfect machinery for the detection of the ISWRS. The results shown in the left panel reach $\sim 19\sigma$, achieved for $M_{\nu} = 0.06$ eV. The promising results for the detection guarantee very promising results also for the identification of the $\nu\Lambda$CDM model, at least for $M_{\nu}>0.18$ eV. 

These findings concretely pave the way for using this probe to infer new constraints on neutrino masses. 
In our forecasts, we neglect any residual foreground contamination in the CMBL reconstruction that might correlate with foreground residuals in the CMB anisotropy map, potentially inducing a systematic bias. However, next-generation CMB experiments will obtain lensing reconstruction primarily from polarization measurements, which are significantly less prone to foreground contamination~\citep{PhysRevD.107.023504, PhysRevD.111.023503}.

\section{Conclusions and discussion}
\label{sec:conclusions}
In this work, we explore the possibility of detecting the ISWRS effect and use this detection to discriminate between $\nu\Lambda$CDM models with current and next generation CMB experiments and LSS surveys. 

 We start by testing an ideal case. We assume to have a $\Phi$-tracer probe whose window function $W(k,z)$ perfectly overlaps with the one of the ISWRS effect (i.e., the investigated signal is actually the ISWRS auto-spectrum, $C_{\ell}^{\dot\Phi\dot\Phi}$), a full-sky coverage and negligible statistical and instrumental noises. We test these conditions using $C_{\ell}^{\dot\Phi\dot\Phi}$ extracted from the DEMNUni N-body simulations realised for the baseline Planck 2014b cosmology~\cite{PlanckXVI_2014}, generalised to $\nu\Lambda$CDM models with $M_{\nu} = 0.0, 0.17, 0.30$ eV. The necessity to work with N-body simulations arises from the impossibility to properly represent nonlinear scales analitically. In this ideal case, not only the ISWRS effect can be detected at $\sim100\sigma$ ($\sim200\sigma$ if delensing techniques are applied to the CMB anisotropy field), but
 it is even possible to discriminate a model with $M_{\nu} = 0.17$ eV from the case of null mass with a $\sim15\sigma$ significance level ($\sim22\sigma$ if delensing techniques are applied), as shown in Figure~\ref{fig:ideal_case}.

After the ideal scenarios, we study optimistic scenarios where we reconstruct GC, CS and CMBL cross spectra assuming full-sky coverage, no detector noise nor foreground residuals in CMB data, a galaxy abundance scenario based on~\cite{Conselice_2016} and an effective galaxy bias model following the halo-bias prescription of~\cite{Tinker_2010}. The results achieved are very promising for each of the cross-correlations and for all the $\nu\Lambda$CDM cosmologies investigated ($M_{\nu} = 0.06, 0.18, 0.30$ eV). In particular, the cross-correlation with CMBL would lead to an ISWRS detection at $\sim40\sigma$, reaching $\sim120\sigma$ if delensing techniques are applied to CMB temperature anisotropy data. In this scenario, GC and CS cross spectra perform similarly, yielding comparable levels of detections that range from $\sim30\sigma$ to $\sim80\sigma$ when delensing is applied. Additionally, the capability to discriminate the $\nu\Lambda$CDM model would reach high statistical significance, even for the $M_{\nu} = 0.06$ eV case when delensing is applied.

Realistic scenarios are definitely less ambitious, yet very promising. Our real-case analysis consists of producing forecasts of the ISWRS cross-correlation with GC, CS and CMBL that will be measured from the combination of SO, CMB-S4, CMB-HD and PICO with Euclid and LSST. To simulate these cross-power spectra, we use our analytical modelling developed in ~\cite{Cuozzo_2024}. We compute them assuming a Planck 2018 baseline cosmology~\cite{PlanckVI_2020}, to which we add neutrinos with total mass $M_{\nu}$ = $0.06$, $0.12$, $0.18$, $0.24$, $0.30$ eV, and using the nonlinear $P_{\delta\delta}(k,z)$ computed with the Takahashi model~\cite{Takahashi_2012}, since we have verified that it works better for the reconstruction of the cross-correlations analysed in our previous work~\cite{Cuozzo_2024}. We test three different scenarios, that account for an increasingly realistic and complete description of the sources of confusion for the measurement, starting from considering only the impact of primary CMB anisotropies, then adding detector noise, and finally also possible residual foregrounds. For the case of the cross-correlation with GC only, we test a fourth scenario, where we apply an optimal weighting to reshape the galaxy distribution to increase the overlap between the ISWRS and the GC window function and minimise the noise. 

We achieve the best results when analysing the cross-correlation of the ISWRS with CMBL, as measured by CMB-HD, across all tested $\nu\Lambda$CDM cosmologies. This holds true for both lensed and delensed $C_{\ell}^{TT}$, and whether considering only detector noise or both detector noise and foregrounds, as evident from Figure~\ref{fig:cmbl}. The reason is that CMB-HD, despite lacking the sky coverage of a space telescope like PICO, exhibits a particularly low CMBL reconstruction noise at the scales of interest due to the high-sensitivity at high-resolution (see Figure~\ref{fig:cmbl_noise}). Considering the extended overlap of the CMBL and ISWRS window functions, the CMBL measured by CMB-HD proves to be a superior probe compared to all the others tested in this work. This is the first time, to our knowledge, that the detectability of the effect has been assessed in cross-correlation with CMBL. Accordingly to its promising detection results, this cross-correlation can also allow the identification of a $\nu\Lambda$CDM model with $M_{\nu} \geq 0.12$ eV, in the case of a measurement with negligible levels of residual foregrounds (current estimates point to $M_{\nu}<0.11$ eV at $95\%$ C.L.;~\cite{Allali_2024}). 

Another implication of the better resolution in the frequency range where CMB dominates (i.e. $70 - 150$ GHz) of CMB-HD, with respect to SO, CMB-S4 and PICO, consists of having lower detector noise and foreground residuals at the scales of interest. These features make CMB-HD the best CMB experiment among those tested in this work even in combination with LSS surveys for the analysis of the other two cross-correlations (Figures~\ref{fig:euclid_gc}-\ref{fig:nbody}). In the case of the cross-correlation with GC, we expect to reach high-significance detections for CMB-S4 and CMB-HD, provided that residual foregrounds can be kept under control. We want to highlight the significant improvement achieved when using the optimal weighting, which leads to an increase of the order of $\sim60\%$ (on average) in the S/N ratio. We obtain the best results for this cross-correlation when combining the LSST ``Optimistic'' sample with CMB-HD, not only for the already mentioned advantages of this CMB experiment, but even for the high galaxy density of this LSS survey (which reduces the galaxy shot-noise), the large overlap between the observed sky fractions of the two instruments (which increases the combined $f_{sky}$), and the significant overlap between the ISWRS and LSST “Optimistic” sample window functions.
Among the GC samples tested, LSST  “Optimistic” is the only one that allows to set constraints on $\nu\Lambda$CDM models with $M_{\nu} \gsim 0.30$ eV, even when accounting for the realistic conditions of lensed $C_{\ell}^{TT}$, with detector noise and residual foregrounds. Moreover, it is the GC sample that reaches the best results when applying the optimal weighting, suggesting the potential of this tool, and in particular of its application to LSST.

The less promising results of this work have been achieved in the analysis of the cross-correlation between the ISWRS effect and the photometric Euclid CS survey. This is not only because the superposition between the window functions of CS and ISWRS effect is significantly lower than the case of GC (see Figure~\ref{fig:windows}) while the noise is of the same order of magnitude, but also because the analytical reconstruction of this cross-correlation is less accurate with respect to the others, and in particular underestimates the results of N-body simulations. These conditions negatively affect our analysis, leading the cumulative S/N ratio to barely reach the $1\sigma$ level (Figure~\ref{fig:euclid_cs}). 

However, we restore confidence in the potential of the cross-correlation between the Euclid CS sample and the ISWRS effect by performing the same analysis with cross-spectra extracted from the DEMNUni N-body simulations (Figure~\ref{fig:nbody}). This test leads to an improved cumulative S/N ratio, enhancing the prospects for detections and restoring the feasibility of disentangling the $\nu\Lambda$CDM models. We believe then that an improvement in the modelling of nonlinear scales is essential to improve the accuracy of our analytical reconstruction and, in particular, to produce more robust predictions of the ISWRS cross-correlation with CS.  

Given that the third scenario (lensed $C_{\ell}^{TT}$ + ILC) we investigate is quite conservative and that, as of today, delensing techniques and foreground cleaning methods are becoming increasingly refined, our predictions are merely a starting point for the promising results that we can expect in the near future. Moreover, considering that even under these conditions we consistently achieve a detection in the case of cross-correlations with GC and CMBL, we feel confident that once experiments similar to those tested here will see the light, we will be able to properly detect the ISWRS. Furthermore, we verify that even  CS is a valid probe for this kind of analysis, despite the technical limitations of our current study. 

Overall, our method is a useful and computationally non-expensive tool to investigate the ISWRS cross-correlations with $\Phi$-tracers in different cosmological scenarios, without the need to run large N-body simulations for each cosmology.

\acknowledgments
VC and MM acknowledge support by the INFN project ``InDark''. MM is also supported by the ASI/LiteBIRD grant n.2020-9-HH.0. MC is partially supported by the 2025/26 ``Research and Education'' grant from Fondazione CRT. MC's institute (OAVdA) is managed by the Fondazione Cl\'ement Fillietroz, which is supported by the Regional Government of the Aosta Valley, the Town Municipality of Nus and the ``Unit\'e des Communes vald\^otaines Mont-\'Emilius''.MC is partially supported by the 2025/26 ``Research and Education'' grant from Fondazione CRT. MC's institute (OAVdA) is managed by the Fondazione Cl\'ement Fillietroz, which is supported by the Regional Government of the Aosta Valley, the Town Municipality of Nus and the ``Unit\'e des Communes vald\^otaines Mont-\'Emilius''. The DEMNUni simulations were carried out in the framework of ``The Dark Energy and Massive-Neutrino Universe'' project, using the Tier-0 IBM BG/Q Fermi machine and the Tier-0 Intel OmniPath Cluster Marconi-A1 of the Centro Interuniversitario del Nord-Est per il Calcolo Elettronico (CINECA). We acknowledge a generous CPU and storage allocation by the Italian Super-Computing Resource Allocation (ISCRA) as well as from the coordination of the ``Accordo Quadro MoU per lo svolgimento di attività congiunta di ricerca Nuove frontiere in Astrofisica: HPC e Data Exploration di nuova generazione'', together with storage from INFN-CNAF and INAF-IA2.

\appendix
\section{Validation of the ISWRS cross-correlation with CS}
\label{app:validation}
\begin{figure*}[!ht]
    \centering
 \includegraphics[width=0.6\textwidth]{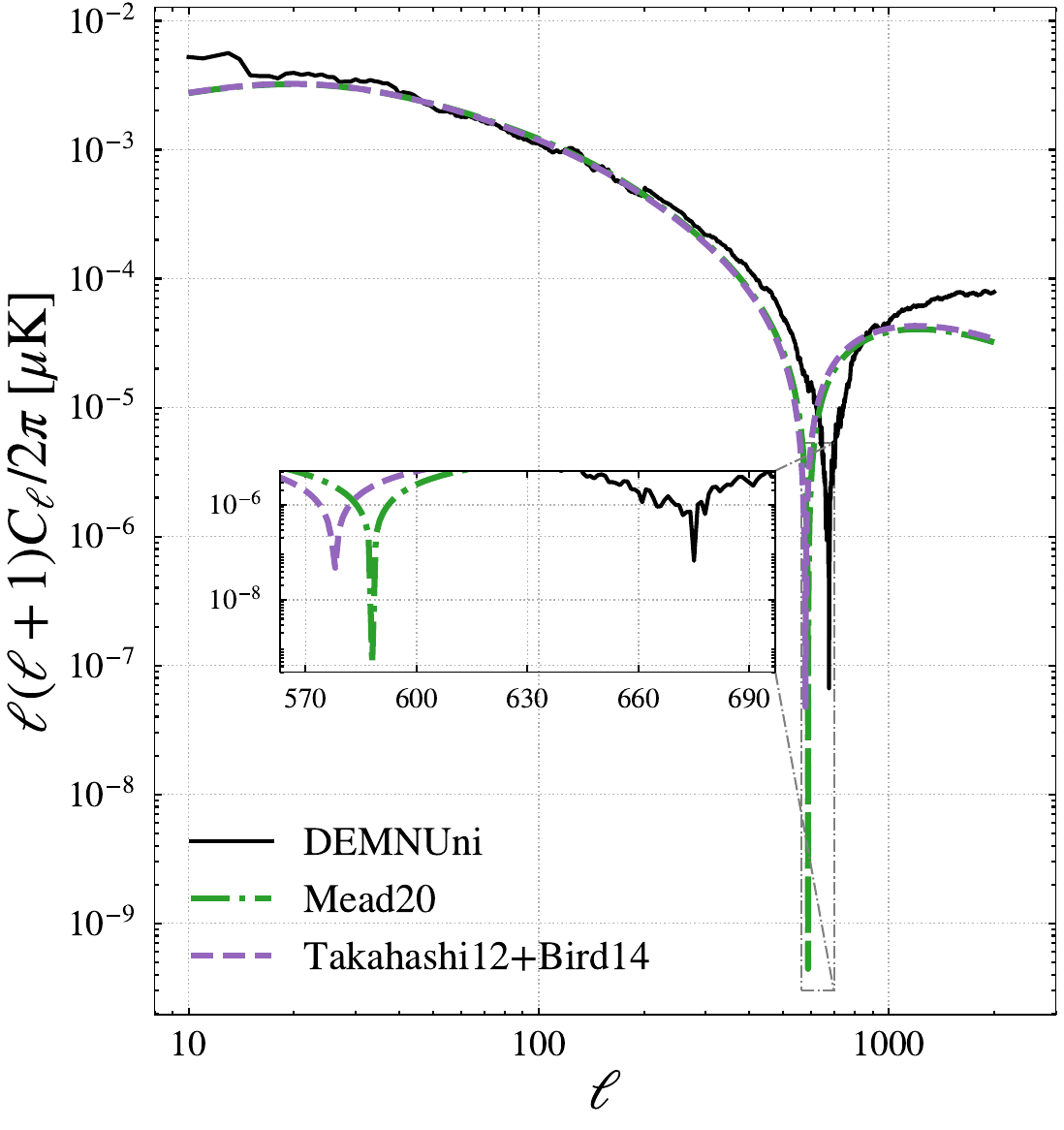}
    \caption{Angular power spectrum obtained from the DEMNUni maps (solid black line), compared with the analytical predictions computed with both the Takahashi12+Bird14~\cite{Takahashi_2012} (dashed violet) and the Mead20~\cite{Mead_2021} (dot-dashed green) \texttt{Halofit} models.}
    \label{fig:cs_validation}
\end{figure*}

Following the procedure implemented in~\cite{Cuozzo_2024}, we validate Equation~\eqref{eq:tw} against DEMNUni simulations. 
We use a set of correlated DEMNUni maps, composed of an ISWRS map and a CS map, realised with the techniques described in~\cite{Carbone_2008, Calabrese_2015, Fabbian_2018}, characterised by the baseline Planck 2014b~\cite{PlanckXVI_2014} cosmology, with $M_{\nu} = 0.0$ eV, in the redshift range that goes from $z_{min} = 0.2$ to $z_{max}=2.5$. We analytically compute the expected cross-angular power spectrum between ISWRS and CS using \texttt{CAMB} for all the ingredients of Equation~\eqref{eq:tw} (assuming the same cosmology as in DEMNUni maps), apart from $n(z)$ given by Equation~\eqref{eq:euclid_n}, which is the galaxy selection used to realise the CS DEMNUni map. For this computation, we test both the Takahashi~\cite{Takahashi_2012} and the Mead2020~\cite{Mead_2021} \texttt{Halofit}~\cite{Smith_2003} modellings. Compared with the cross-power spectrum extracted from the DEMNUni maps, the overall shape of the spectrum and the amplitude at very large scales are accurately recovered (see Figure~\ref{fig:cs_validation}). However, differently from the results presented in~\cite{Cuozzo_2024}, where the  cross-correlations of ISWRS with GC and CMBL were validated, the discrepancies in the position of the sign inversion and the amplitude at small scales are more pronounced. This is likely because the CS field is more sensitive to nonlinearities, and the assumptions made in the considered \texttt{Halofit} models do not offer precise reconstructions of nonlinear scales for this cross-correlation. This mismatch at small scales is responsible for our discouraging results of Section~\ref{sec:cross-cs}. 

\section{Cross angular power spectra}
\label{app:ccs}
The cross angular power spectrum between two fields \textbf{A} and \textbf{B} can be written as~\cite{Planck_XX1_2016, Smith_2009}:
\begin{equation}
    C^{AB}_{\ell} = \frac{2}{\pi} \, \int {\rm d}k\,  \int {\rm d}z\,  \int {\rm d}z'\,  k^{2} \, P_{\delta\delta}(k,z,z') \, W^{A}(k,z) \,j_{\ell}[k\,\chi(z)] \, W^{B}(k,z') \,j_{\ell}[k\,\chi(z')] \,,
\label{eq:cl_AB}
\end{equation}
where $j_{\ell}[k\,\chi(z)]$ are the Bessel functions and $W(k,z)$ are the window functions of the selected fields. The Limber approximation allows us to have~\cite{Limber_53, Smith_2009}:
\begin{equation*}
    \int {\rm d}k\, k^{2} \, P_{\delta\delta}(k,z,z')\,j_{\ell}[k\,\chi(z)]\,j_{\ell}[k\,\chi(z')] \simeq  \frac{\pi}{2} \, k^{2} \, \delta_{D}(z - z') \, \frac{H(z)}{c} \, \frac{1}{k^{2}\, \chi^{2}(z)} P_{\delta\delta}\left(k=\frac{\ell+1/2}{\chi(z)},z\right) \,
\end{equation*}
where we have used:
\begin{equation*}
    \delta_{D}[\chi(z) - \chi(z')] = \delta_{D}(z - z') \, \frac{H(z)}{c} \,,
\end{equation*}
in the Dirac delta, given that:
\begin{equation*}
    \chi(z) = \int {\rm d}z' \, \frac{c}{H(z')} \,.
\end{equation*}
Fixing now:
\begin{equation}
    k=\frac{\ell+1/2}{\chi(z)} \,,
\end{equation}
equation~\eqref{eq:cl_AB} becomes then:
\begin{equation}
C^{AB}_{\ell} =  \int {\rm d}z\, \frac{H(z)}{c\, \chi^{2}(z)} \, W^{A}(k,z)  \, W^{B}(k,z) P_{\delta\delta}\left(k,z\right)  \,.
\label{eq:cl_AB_final}
\end{equation}

However, in the case of the cross-correlation between the ISWRS effect ($\dot\Phi$) and a gravitational potential tracer \textbf{Y}, we have to compute:
\begin{equation}
C^{\dot\Phi \text{Y}}_{\ell} =  -2 \int {\rm d}z\, \frac{H(z)}{c\, \chi^{2}(z)} \, W_{\text{Y}}(z) P_{\dot\Phi,\delta}(k,z)  \,,
\label{eq:cl_iY_final}
\end{equation}
where the factor $(-2)$ comes from Equation~\eqref{eq:iswrs} (neglecting $e^{- \tau(z)}$~\cite{Stolzner_2018}).

Starting from the Poisson's equation~\cite{Cai_2010} expressed as a function of $t$:
\begin{equation*}
    \Phi(k,t) = - \frac{3}{2} \, \Omega_{m} \, \left( \frac{H_{0}}{c\,k}\right)^{2} \, \frac{\delta(k,t)}{a(t)} \,,
\end{equation*}
one can derive:
\begin{equation*}
    \delta(k,t) = -\frac{a(t)}{F(k)} \Phi(k,t) \,,
\end{equation*}
with 
\begin{equation}
    F(k) = \frac{3}{2} \, \Omega_{m}\, \left( \frac{H_{0}}{c\,k}\right)^{2}\,.
\end{equation}

We have then~\cite{Smith_2009}:
\begin{align*} 
    P_{\delta,\dot\Phi}(k,t) &= \, \left<\delta(k,t), \dot\Phi(k,t)\right> =  \\\nonumber
    &= -\frac{a(t)}{F(k)} \, <\Phi(k,t), \dot\Phi(k,t)> =  \\\nonumber
    &= -\frac{a(t)}{F(k)} \, P_{\Phi,\dot\Phi}(k,t) = \\\nonumber
    &\simeq -\frac{a(t)}{F(k)} \, \frac{1}{2} \partial_{t}[P_{\Phi,\Phi}(k,t)] = \\\nonumber
     &\simeq - a(t) \, F(k) \, \frac{1}{2} \partial_{t}\left[\frac{P_{\delta,\delta}(k,t)}{a^{2}(t)}\right] \,
\end{align*}

where we use $P_{\Phi,\Phi}(k,t) = \left(\frac{F(k)}{a(t)}\right)^{2}\, P_{\delta,\delta}(k,t)$ and $P_{\Phi,\dot\Phi}(k,t)\simeq \frac{1}{2} \partial_{t}[P_{\Phi,\Phi}(k,t)]$~\cite{Smith_2009, Nishizawa_2008}.

Making now a coordinates transformation from $t$ to $z$, we have:
\begin{equation*}
    (-2)\, W_{\text{Y}}(z) P_{\delta,\dot\Phi}(k,z) = \frac{3}{2} \Omega_{m} \, \left(\frac{H_{0}}{c \,k}\right)^{2} \, a(z) \, W_{Y}(z) \, \left[\partial_{z} \frac{P_{\delta\delta}(k,z)}{a^{2}(z)}\right] \,, 
\end{equation*}
that with the proper $W_{\text{Y}}(z)$ from Equations~\eqref{eq:windows} results in Equations~\eqref{eq:cross_cls}.

\bibliographystyle{JHEP}
\bibliography{bibliography.bib}

\appendix

\end{document}